\documentclass[twocolumn]{aastex631}

\usepackage[T1]{fontenc}
\usepackage[utf8]{inputenc}
\usepackage{graphicx}
\usepackage{amssymb}
\usepackage{amsmath}
\usepackage{natbib}
\usepackage[capitalise]{cleveref}
\usepackage{booktabs}
\usepackage{bm}
\usepackage{comment}
\usepackage{gensymb}
\usepackage{enumitem}
\usepackage[dvipsnames]{xcolor}

\definecolor{ncsgreen}{rgb}{0.2, 0.5, 0.0}

\newcommand{\blue}[1]{#1}
\newcommand{\red}[1]{}

\begin{document}

\title{EMRIs in Kerr Spacetime: The Role of Inclination Diffusion}
\correspondingauthor{Yael Alush}
\author[0000-0003-3471-9817]{Yael Alush}
\email{yael.alush@mail.huji.ac.il}
\affiliation{Racah Institute of Physics, The Hebrew University, Jerusalem, 91904, Israel}
\affiliation{Department of Astronomy, University of Wisconsin, Madison, WI, 53706}
\author[0000-0002-4337-9458]{Nicholas C.~Stone}
\affiliation{Racah Institute of Physics, The Hebrew University, Jerusalem, 91904, Israel}
\affiliation{Department of Astronomy, University of Wisconsin, Madison, WI, 53706}
\author{Scott A. Hughes}
\affiliation{Department of Physics and MIT Kavli Institute, MIT, Cambridge, MA 02139 USA}

\begin{abstract}
Extreme mass ratio inspirals (EMRIs) are expected to be a major source of gravitational waves (GWs) for the future {\it Laser Interferometer Space Antenna} ({\it LISA}). An EMRI occurs when a compact object, driven by the emission of gravitational radiation, slowly inspirals into a massive black hole (MBH). The classical formation channel involves repeated two-body interactions within dense nuclear star clusters. However, both EMRI event rates and the expected distribution of their orbital parameters remain highly uncertain. A major source of this uncertainty is the poorly constrained ratio between gradually inspiraling EMRIs and direct plunges, which do not produce detectable GWs. While most studies of astrophysical EMRI production focus on the Schwarzschild limit, previous work examining the Kerr metric found that the spin of the central MBH can significantly affect the EMRI-to-plunge ratio. This enhancement depends sensitively on orbital inclination, yet earlier studies held inclination fixed. In this work, we develop the mathematical framework describing inclination evolution driven by two-body encounters evolving along Kerr geodesics. By implementing 3D diffusion into Monte Carlo simulations, we show that initially prograde orbits may evolve into retrograde states, thereby converting potential EMRIs into plunges or vice versa. We estimate the EMRI-to-plunge ratio and find that substantial MBH spin enhances the EMRI rate by a factor of $\sim 2.5$. Additionally, even for a non-spinning MBH, modeling the orbits as relativistic geodesics rather than Newtonian trajectories increases the number of EMRIs by a similar factor of $\sim 2.5$. The distribution of Kerr EMRI inclinations has a substantial prograde bias that may aid in identifying the subset of loss cone EMRIs within large future EMRI samples.

\end{abstract}

\section{Introduction}
\label{sec:intro}

The first discoveries of Advanced LIGO inaugurated the era of gravitational wave (GW) astronomy \citep{Abbott+2016a}, opening a window on previously unobserved black hole (BH) populations \citep{Abbott+23} and enabling novel strong-field tests \citep{Abbott+2016a} of general relativity (GR).  In the 2030s, the launch of space-based interferometers such as {\it LISA} and {\it TianQin/Taiji} will extend this window to lower frequencies \citep{Danzmann+1996, Luo+2016, Amaro-Seoane2018} sensitive to massive black holes (MBHs).  MBH-MBH mergers will provide the highest signal-to-noise ratios (SNRs in the thousands are likely; \citealt{Valiante+2021}) out of all GWs seen in the {\it LISA} band, but it may ultimately be the slower inspiral of a stellar mass compact object (CO) into a nuclear\footnote{If intermediate mass black holes are common in globular clusters, these systems may also be important sources of EMRIs \citep{GonzalezPrieto+25}.} MBH that provides the strongest tests of GR.  These extreme mass ratio inspirals (EMRIs; \citealt{Ryan1995}) will stay in the {\it LISA} band for up to $\sim 10^5$ cycles \citep{Babak2017}, accumulating an unprecedented degree of phase and timing information and therefore offering exquisite sensitivity to small deviations from GR \citep{Yunes+2012, Speri+2024} or the Kerr metric \citep{Ryan1997, CollinsHughes2004}; see \citet{Barausse+2020} for a broad review.  The astrophysics of EMRIs, however, remains highly uncertain.

Because classical EMRIs are electromagnetically dark\footnote{The recent discovery of quasi-periodic eruptions (QPEs; \citealt{Miniutti+2019, Giustini+2020, Arcodia+2021}) from a range of galactic nuclei is interesting in this regard, as a leading theory is that these X-ray outbursts are triggered by collisions between an EMRI and an MBH accretion disk \citep{Dai+2010}.  While these may be true EMRIs involving a CO \citep{Franchini+2023}, it is more likely that they are GW inspirals of main sequence stars \citep{Linial&Metzger2023, Tagawa&Haiman2023}.  Regardless, QPEs may prove useful for calibrating true EMRI rates.}, present-day estimates for their event rates and parameters are ultimately theoretical and subject to large uncertainties.  
Despite some Milky Way constraints \citep{Hailey+2018, Generozov+2018}, the population of nuclear neutron stars (NSs) and stellar mass black holes (BHs) is largely unknown outside our Galaxy.  
More fundamentally, it is currently unclear what is the dominant mechanism for feeding COs to MBHs.  Most past work on this subject has focused on scatterings into the MBH loss cone \citep{FrankRees1976, CohnKulsrud1978}: a conical region of velocity space\footnote{Equivalently, a parabaloidal region of angular momentum space.} inside of which a test CO's pericenter falls inside a critical value that would trigger its capture by the MBH \citep{HilsBender1995, SigurdssonRees1997}.  Stochastic scatterings with field stars will lead COs to undergo a random walk in angular momentum, eventually pushing their pericenters inside this loss cone.  If the CO is tightly bound to the MBH, the relevant loss cone is set by GW inspiral timescales and results in EMRI production; if the CO is more loosely bound, GW timescales are too long and the CO will plunge directly and unobservably into the MBH event horizon \citep{Hopman&Alexander2005}.

EMRI rates even in this well-studied scenario are uncertain by at least three orders of magnitude \citep{Babak2017, Broggi+2022}, with major systematic uncertainties stemming from the plunge-to-EMRI ratio \citep{QunbarStone2024, Mancieri+2025}, the distribution of low-mass (``intermediate mass'') MBHs in dwarf galactic nuclei \citep{Greene+2020}, and the ability of stellar mass BHs to replenish themselves through dynamical friction \citep{Rom+2024} or {\it in situ} formation \citep{AharonPerets2016}.  Other EMRI formation scenarios that have been proposed more recently are subject to even larger and less well-characterized uncertainties. 
These scenarios include GW-driven inspirals of BHs deposited by the Hills mechanism \citep{Miller+2005, RavehPerets2021}; Kozai-Lidov cycles in MBH-MBH binaries \citep{Bode&Wegg2014,Naoz+2022, Mazzolari+2022,Naoz&Haiman2023}; and gas-assisted inspirals in active galactic nuclei (AGN; \citealt{Levin2007, Pan+2021, DerdzinskiMayer2023}).  

Given the present state of theoretical EMRI predictions, it may seem unkind to draw attention to another systematic uncertainty, but that is the purpose of this paper.  A key variable strongly influencing both volumetric EMRI rates and peak GW strain is the spin, $\chi_\bullet$, of the primary MBH.  Larger $\chi_\bullet$ allows EMRIs to approach closer to the horizon before plunging, producing stronger signals detectable over a larger cosmological volume.  More subtly, $\chi_\bullet$ can dramatically affect the competition between GW dissipation and stochastic scattering that determines the EMRI-to-plunge ratio in loss cone channels.  \citet{AmaroSeoane+2013} investigated this competition in the limit of fixed inclination, and found that as $\chi_\bullet$ increases from 0 to values near 1, the EMRI rate can in principle increase by a factor of $\approx 30$, due to the larger phase-space volume from which EMRIs can form, thanks to the enhancement in per-orbit GW dissipation for prograde, near-equatorial orbits around MBHs of increasing spin.
Recent work continuing to operate in the fixed-inclination approximation has likewise found systematically higher event rates around spinning MBHs, but with a total rate enhancement that is modest when averaged over inclinations \citep{ZhangAmaroSeoane2026}.

In this paper, we re-examine the role of MBH spin in determining EMRI event rates and orbital parameter distributions.  We go beyond the fixed inclination approximation by developing the mathematical framework of inclination diffusion from two-body scatterings\footnote{As this paper was in its final stages, a different preprint focused on the tidal disruption event loss cone was released \citep{Xin2026}; a more detailed discussion of similarities and differences is in \S \ref{sec:discuss}.}. Building on these generalized diffusion coefficients, we perform Monte Carlo simulations of individual BH trajectories as they diffuse through three-dimensional action space around a central {\it spinning} MBH.  
In \S \ref{sec:GDC}, we take the standard formalism for stellar diffusion via two-body scatterings, normally applied to Keplerian orbits, and extend it into Kerr metric geodesics.  In \S \ref{sec:GW}, we review an analytic kludge scheme to estimate GW losses and their impact on the Kerr integrals of motion, and compare these fluxes to Teukolsky-based solutions. In \S \ref{sec:algo}, we present our numerical algorithm for solving the Langevin-like equation governing the orbit-averaged evolution of test BHs in Kerr spacetime. In \S \ref{sec:results}, we present results from both individual Monte Carlo simulations and large statistical ensembles of these simulations.
In \S \ref{sec:discuss}, we discuss the astrophysical implications, 
and in \S \ref{sec:conc} we summarize our main conclusions.

\section{Geodesic Diffusion Coefficients}
\label{sec:GDC}

Dense stellar environments such as galactic nuclei can be understood as weakly collisional kinetic systems.  Weak, uncorrelated two-body scatterings between stars will stochastically perturb the velocity vector of a test star (or BH).  By integrating over distributions of field star velocities and impact parameters, one can derive velocity-space diffusion coefficients (DCs) encapsulating this process.  This statistical formalism was first pursued to study dynamical friction \citep{Chandrasekhar43} and globular clusters \citep{SpitzerHart1971}; for a thorough review of stellar diffusion in galactic nuclei, see \citet{Merritt2013}.  As galactic nuclei are both quasi-spherical and dominated by the gravitational field of the central MBH, stellar orbits are integrable, or nearly so (ignoring for now GW dissipation), and it can be more fruitful to work with DCs in action space.  

Many previous works have studied EMRI formation in the loss cone channel via these types of action-space DCs \citep{Hopman&Alexander2005, Bar-Or&Alexander2016, RavehPerets2021, Broggi+2022, QunbarStone2024, Mancieri+2025}.  Because orbits typically evolve only slowly\footnote{Though see \citet{KaurPerets2024} for the role that strong scatterings can play in pre-emptively ejecting BHs before an EMRI.}, these DCs are often orbit-averaged to simplify the problem.  Orbit-averaged DCs can be used to find semi-analytic or analytic solutions to the Fokker-Planck equation \citep{LightmanShapiro1977}, describing the evolution of large populations of test objects.  Alternatively, they can be applied on a particle-by-particle basis to solve a Langevin-like stochastic differential equation \citep{ShapiroMarchant1978}, as we will do in this work.  Past approaches usually assumed spherical symmetry (both in the gravitational potential and the distribution of field stars) so that diffusion can be studied in just two integrals of motion, orbital energy $E$ and total angular momentum $L$.  

In this paper we are especially concerned with the role played by inclination diffusion for geodesics in the axisymmetric Kerr metric, which has three integrals of motion: $E$, the $z$-component of angular momentum $L_z$, and the Carter constant $Q$ \citep{Bardeen+1972}.  In this section, we therefore derive an extension to the classic 2D theory of stellar diffusion into the 3D space of Kerr geodesic integrals of motion, with the ultimate goal of producing geodesic diffusion coefficients (GDCs).  

\subsection{Approximate Theory of EMRI Formation}
\label{sec:twiddle}


When a CO of mass $m_\bullet$ orbits around a MBH of mass $M_\bullet$ on a highly eccentric orbit with semimajor axis $a$ and eccentricity $e$, two characteristic timescales govern the evolution toward an EMRI: 
\begin{itemize}
    \item The angular momentum relaxation timescale, $t_{\rm rlx}$, defined as the timescale over which two-body encounters induce significant changes in the orbital parameters.
    \item The GW inspiral timescale, $t_{\rm GW}$, defined as the timescale over which GW emission drives significant orbital evolution.
\end{itemize}

At large distances from the MBH, GW emission is negligible, and the orbital evolution is dominated by stochastic two-body encounters with other stars in the cluster. These encounters alter both the orbital energy and angular momentum (equivalently, the semimajor axis and eccentricity), but are generally more efficient at changing angular momentum than energy \citep{Alexander2017}. 

To model the timescale of this angular momentum diffusion, we consider a simplified description of the stellar background. We assume a spherically symmetric stellar cluster with a power-law density profile, $\rho(r)=\rho_{\rm inf}(r/r_{\rm inf})^{-\gamma}$, where $r_{\rm inf}$ is the influence radius and $\rho_{\rm inf}$ is the stellar mass density at that radius. The slope $\gamma$ characterizes the steepness of the stellar cusp (e.g., $\gamma=7/4$ corresponding to a relaxed \citealt{Bahcall&Wolf1976} cusp).  We further assume a Keplerian potential, $\Phi=-GM_\bullet/r$, such that the local velocity dispersion is $\sigma^2(r)=GM_\bullet/[r(1+\gamma)]$. Under these assumptions, the angular-momentum relaxation timescale can be estimated as \citep{Hopman&Alexander2005,BortolasMapelli19}:
\begin{equation}
    t_{\rm rlx}\sim \frac{L^2}{\langle \dot{L}^2 \rangle}\sim \frac{2k\sigma^3(r)}{G^2n(r)m^2\ln{\Lambda}}(1-e^2)
    \label{eq:AM relaxation timescale}
\end{equation}
where $k\approx0.34$ \citep{Binney&Tremaine2008}, $n(r)=\rho(r)/m$ is the stellar number density, $\ln{\Lambda}\approx\ln{(0.4M_\bullet/m)}$ is the Coulomb logarithm, $m$ is the characteristic stellar mass in the cluster, and $\langle \dot{L}^2 \rangle$ is a time-averaged rate of (squared) angular momentum evolution due to stochastic scatterings.  

As the CO undergoes stochastic evolution in energy and angular momentum, it can reaches close enough to the MBH that GW emission becomes important. At each pericenter passage, the orbit loses energy and angular momentum, which causes both the semimajor axis and the eccentricity to decrease. The orbit-averaged specific energy and angular momentum losses at leading order are \citep{Peters1964}:
\begin{equation}
    \begin{aligned}
        \Delta E_{\rm GW}&= -\frac{64\pi G^{7/2}M_\bullet^{5/2}m_\bullet}{5 c^5 a^{7/2} (1-e^2)^{7/2}} \left(1+\frac{73}{24} e^2+\frac{37}{96}e^4\right),\\
        \Delta L_{\rm GW}&= -\frac{64\pi G^3 M_\bullet^2 m_\bullet}{5c^5 a^2 (1-e^2)^2}\left(1 + \frac{7}{8} e^2\right).
    \end{aligned}
\end{equation}
For a highly eccentric orbit, the GW inspiral timescale is \citep{Peters1964}:
 \begin{equation}
    t_{\rm GW}=\frac{24\sqrt{2}}{85}\frac{a^4c^5}{ G^3 M_\bullet^2 m_\bullet}(1-e)^{7/2}.
    \label{eq: GW timescale}
 \end{equation}

In the picture described above, where the orbital evolution is governed either by two-body scatterings or by GW dissipation, the relative size of the two timescales determines the evolution of the CO orbit. When $t_{\rm rlx}<t_{\rm GW}$, the CO evolves stochastically and two-body scatterings dominate the changes in its orbital parameters. When $t_{\rm GW}<t_{\rm rlx}$, GWs take over and the orbit evolves deterministically, shrinking and circularizing as the CO approaches the MBH. Once $t_{\rm GW}\ll t_{\rm rlx}$ an EMRI forms and the CO undergoes a slow inspiral.

Not all orbital evolutions produce long-lived GW signals. If the CO’s pericenter enters the loss cone, the object plunges directly through the MBH event horizon and does not generate a detectable inspiral. Previous studies show that such plunging orbits emit only a short GW burst, and are likely to be detectable only for events originating in the Galactic Center \citep{Hopman+2007}. These orbits are therefore considered unobservably lost from the system. The boundary defining the Schwarzschild loss cone is the innermost bound circular orbit (IBCO), corresponding to the last parabolic pericenter \citep{Bardeen+1972} with angular momentum is $L_{\rm IBCO}=4GM_\bullet/c$. In the Kerr case, however, the IBCO generalizes to the innermost bound spherical orbit (IBSO), depends on the orbital inclination, and can be either smaller or larger than the Schwarzschild IBCO. Since the IBSO determines the threshold between inspiral and direct plunge, it plays a central role in setting the EMRI-to-plunge ratio.

Because $t_{\rm GW}$ decreases more rapidly than $t_{\rm rlx}$ as the CO approaches the IBSO, there exists a critical semimajor axis that separates true EMRIs from plunges \citep{Hopman&Alexander2005}. This critical value is obtained by equating $t_{\rm rlx}= t_{\rm GW}$ and requiring that the resulting pericenter be set to the IBSO:
\begin{equation}
    a_{\rm c}^{3-\gamma}= \frac{85\pi}{6 \sqrt{2}}\frac{k}{(1 + \gamma)^{3/2}(3 - \gamma)}\frac{G^{5/2} M_\bullet^{5/2}m_\bullet r_{\rm inf}^{3 - \gamma}}{c^5 r_{\rm IBSO}^{5/2} m\ln{\Lambda}}.
\end{equation}
An orbit starting with an initial semimajor axis below the critical value, $a\lesssim  a_{\rm c}$, will eventually reach the region where $t_{\rm GW}< t_{\rm rlx}$ before crossing the IBCO, and will therefore become an EMRI. In contrast, if the initial semimajor axis is larger than the critical value, $a\gtrsim a_{\rm c}$, two-body scatterings always dominate and the orbit will eventually plunge into the MBH.

Previous research shows that this simple EMRI-plunge dichotomy does not hold for intermediate-mass black holes and more massive COs \citep{QunbarStone2024,Mancieri+2025}. In these systems, strong GW energy dissipation over a single orbital period can drive objects with semimajor axes $a>a_c$ to $a<a_c$. As a result, objects that would traditionally be classified as direct plunges can instead transition into GW-driven inspirals and form EMRIs. This population of EMRIs is referred to as ``cliffhanger'' EMRIs.

This classic picture of a single critical semimajor axis is further modified in the presence of MBH spin \citep{AmaroSeoane+2013,Vazquez-Aceves+2022,Sun+2026, ZhangAmaroSeoane2026}. 
The location of the Kerr IBSO depends on both the MBH spin and the orbital inclination. In particular, relative to the non-spinning case, the IBSO radius decreases for prograde orbits and increases for retrograde orbits \citep{Stein&Warburton2020} . This leads to an enhanced EMRI fraction for prograde orbits and a higher plunge fraction for retrograde ones. Because the IBSO is not symmetric between prograde and retrograde orbits, the net effect is an overall excess of EMRIs over plunges. Consequently, a spinning MBH is characterized not by a single IBCO and corresponding critical semimajor axis, but by a range of IBSOs that define a corresponding range of critical semimajor axes.

To estimate how the MBH spin, and therefore the orbital inclination, affect the EMRI rates, \cite{AmaroSeoane+2013} define the ratio between the IBSO radii in the Kerr and the Schwarzschild cases:
\begin{equation}
    \mathcal{W}(\chi_\bullet, \iota)\equiv \frac{r_{\rm IBSO, Kerr}}{r_{\rm IBSO, Schw}}. 
\end{equation}
Using this ratio, they calculated the corresponding change in the \blue{critical semimajor axis and the} EMRI rate:
\blue{
\begin{equation}
    a_{\rm c}^{\rm Kerr}=a_{\rm c}^{\rm Schw}\mathcal{W}^{\frac{-5}{6-2\gamma}}(\chi_\bullet,\iota),
\label{eq: fixed I ac}
\end{equation}
}
\begin{equation}
    \dot{N}_{\rm EMRI}^{\rm Kerr}=\dot{N}_{\rm EMRI}^{\rm Schw}\mathcal{W}^{\frac{20\gamma-45}{12-4\gamma}}(\chi_\bullet,\iota).
\label{eq: fixed I rate}
\end{equation}
As an example, for a spin of $\chi_\bullet = 0.999$ and an inclination of $\iota = 0.4$ rad, the enhancement of the EMRI rate relative to the Schwarzschild case depends on $\gamma$. Specifically, the rate increases by a factor of about $67$ for $\gamma = 1$, $30$ for $\gamma = 3/2$, $15$ for $\gamma = 7/4$, and $5$ for $\gamma = 2$. This illustrates that while rapid black hole spin can significantly boost EMRI rates, the magnitude of this enhancement decreases for steeper density profiles. 

However, this estimate assumes that the inclination remains fixed during the evolution. While GW emission induces only negligible changes in the inclination during inspiral \citep{Hughes:2001jr}, two-body scatterings can significantly change a system's inclination. The inclination relaxation timescale is estimated in Appendix \ref{appendix: inclination timescale} and is related to the angular-momentum relaxation timescale (in Eq. \ref{eq:AM relaxation timescale}) as:
\begin{equation}
   t_\iota\sim \frac{\sin^2{\iota}}{\sin^2{\iota}-\cos^2\theta}t_{\rm rlx}.
\end{equation}
Orbit-averaging over $\theta$ results in 
\begin{equation}
    t_\iota \sim 2 t_{\rm rlx}.
    \label{eq: inclination timescale}
\end{equation}
To investigate the impact of inclination evolution on EMRI rates in the presence of MBH spin, it is therefore necessary to extend the classical 2D Newtonian framework of stellar diffusion coefficients to the full 3D Kerr spacetime along geodesic motion. This mathematical framework is developed in the next section. 

\subsection{Local Diffusion Coefficients}
\label{sec:local}

To describe the effect of \red{weak} gravitational encounters, we use the Fokker–Planck equation. It describes the change in the distribution function produced by collisions, $\left(\partial f_b/\partial t\right)_c$, where $f_b(t,\Vec{x},\Vec{v})$ is the phase-space number density of field stars of mass $m_b$, depending on time $t$, position $\Vec{x}$, and velocity $\Vec{v}$. Assuming that velocity changes $\Delta v^i$ are small over short time intervals, higher-order terms can be neglected, yielding
\begin{equation}
    \left(\frac{\partial f_b}{\partial t}\right)_c=-\frac{\partial}{\partial v^i}\left(f_b\langle\Delta v^i\rangle\right)+\frac{1}{2}\frac{\partial^2}{\partial v^i\partial v^j}\left(f_b\langle\Delta v^i\Delta v^j\rangle\right)
\end{equation}
where $v^i$ ($i=1,2,3$) is the $i$th component of velocity in Cartesian coordinates and $\langle\Delta v^i\rangle$ is the average change in $v^i$ per unit time experienced by a test star through interactions with all other stars in the cluster. In this section, we adopt the Einstein summation convention. 

\red{The derivation of the Fokker-Planck equation assumes that only distant encounters are relevant, so that changes in $v^i$ are small. Accordingly, only the first-order $\langle\Delta v^i\rangle$ and second-order  $\langle\Delta v^i\Delta v^j\rangle$ diffusion coefficients are considered. The diffusion coefficients for a test object of mass $m$, are given by } \blue{Under the assumption that weak, distant encounters dominate the scatterings, the diffusion coefficients for a test object of mass $m_\bullet$, are given by} 
\begin{equation}
\begin{aligned}
\langle\Delta v^i\rangle&=\Gamma'\sum_bm_b(m_\bullet+m_b)\frac{\partial h_b}{\partial v^i},\\
\langle\Delta v^i\Delta v^j\rangle&=\Gamma'\sum_b m_b^2\frac{\partial^2 g_b}{\partial v^i \partial v^j},\\
h_b(\Vec{v})&\equiv \int\frac{f_b(\Vec{v}_f)}{|\Vec{v}-\Vec{v}_f|}{\rm d}\Vec{v}_f\\
g_b(\Vec{v})&\equiv \int f_b(\Vec{v}_f) |\Vec{v}-\Vec{v}_f| {\rm d}\Vec{v}_f
\end{aligned}
\end{equation}
where $\Gamma'\equiv4\pi G^2\ln{\Lambda}$, $G$ is the gravitational constant, and $\ln{\Lambda}\approx\ln{(0.4M_\bullet/m_\bullet)}$ is the Coulomb logarithm. The sums over $b$ run over all stellar species in the system, each with mass $m_b$ and distribution function $f_b$. 

The diffusion coefficients above are expressed in Cartesian velocity components.  Our goal is to work in terms of the integrals of motion of the system. In the Kerr metric, there are three such integrals: the energy $E$, the $z$-component of the angular momentum $L_z$ and the Carter constant $Q$. Two-body relaxation is a local process that predominantly occurs in the weak-field regime, where relativistic corrections to individual scatterings are negligible. This allows us to approximate the integrals of motion by their Newtonian expressions when computing the diffusion coefficients. In Boyer-Lindquist coordinates $(r,\theta,\phi)$, which reduce to spherical coordinates far from the BH, the specific values of these constants can be written as
\begin{subequations}
    \begin{align}
        E&=c^2+\frac{v^2}{2}+\Phi(\Vec{x})\\
        L_z&=r\sin{\theta}v_\phi\\
        Q&=r^2v_\theta^2+\cos^2\theta\left[\chi_\bullet^2\left(1-E^2/c^4\right)+L_z^2/\sin^2\theta\right]
    \end{align}
    \label{eq: consts of motion}
\end{subequations}
where $0\leq \chi_\bullet\leq GM_\bullet/c$ is the spin of the MBH with mass $M_\bullet$, and $\Phi(\Vec{x})$ is the \red{stellar potential} \blue{gravitational potential}. 

In Appendix \ref{appendix sec: GDC}, we derive the local diffusion coefficients for the three integrals of motion in \cref{eq: consts of motion} for a general, non-spherical field of stars. The general expressions are given in \cref{appendix eq: general LDC,appendix eq: general LDC cross terms}. While these expressions apply to arbitrary velocity distributions, in this paper we focus on the simpler case of a potential which depends only on the radial coordinate $r$, $\Phi(r)$ (i.e., one which asymptotes to spherical symmetry far from the BH), and with an isotropic field-star velocity distribution. In such a system, $h_b$ and $g_b$ depend only on $v$, which significantly simplifies the diffusion coefficients. In velocity space, the only non-zero diffusion coefficients up to second order are: 
\begin{equation}
\begin{aligned}
\langle\Delta v_\parallel\rangle&=\Gamma'\sum_bm_b(m+m_b)\frac{dh_b}{dv},\\
\langle(\Delta v_\parallel)^2\rangle&=\Gamma'\sum_bm_b^2\frac{d^2g_b}{dv^2},\\
\langle(\Delta v_\perp)^2\rangle&=\frac{2\Gamma'}{v}\sum_bm_b^2\frac{dg_b}{dv}
\end{aligned}
\end{equation}
where $v_\parallel$ is the velocity component parallel to the motion of the test object, and $v_\perp$ is the component perpendicular to it. In the isotropic limit, $h_b$ and $g_b$ reduce to: 
\begin{equation}
\begin{aligned}
    h_b(v)&=4\pi v[I_2(v)+J_1(v)],\\
    g_b(v)&=\frac{4\pi v^3}{3}[3I_2(v)+I_4(v)+3J_3(v)+J_1(v)],\\
    I_n(v)&\equiv\int_0^v\left(\frac{v_f}{v}\right)^nf_b(v_f)dv_f,\\
    J_n(v)&\equiv\int_v^\infty\left(\frac{v_f}{v}\right)^nf_b(v_f)dv_f.
\end{aligned}
\end{equation}

Additionally, two-body scattering is most effective near the apocenter, far from the MBH, where the influence of the spin $\chi_\bullet$ is negligible. Consequently, spin-dependent terms can be omitted from the coefficients. With these simplifications, the local diffusion coefficients reduce to:
\begin{equation}
\begin{aligned}
    \langle\Delta E\rangle&=\Gamma'\left[v\frac{\partial\mathcal{H}}{\partial v}+\frac{1}{2}\frac{\partial^2\mathcal{G}}{\partial v^2}+\frac{1}{v}\frac{\partial\mathcal{G}}{\partial v}\right]\\
    &=v\langle\Delta v_\parallel\rangle+\frac{1}{2}\langle(\Delta v_\parallel)^2\rangle+\frac{1}{2}\langle(\Delta v_\perp)^2\rangle,\\
    \langle(\Delta E)^2\rangle&=\Gamma'\left[v^2\frac{\partial^2\mathcal{G}}{\partial v^2}\right]=v^2\langle(\Delta v_\parallel)^2\rangle,\\
    \langle\Delta L_z\rangle&=\Gamma'\left[\frac{L_z}{v}\frac{\partial\mathcal{H}}{\partial v}\right]=\frac{L_z}{v}\langle\Delta v_\parallel\rangle,\\
    \langle(\Delta L_z)^2\rangle&=\Gamma'\left[\left(\frac{L_z}{v}\right)^2\frac{\partial^2\mathcal{G}}{\partial v^2}+\frac{r^2\sin^2\theta v^2-L_z^2}{v^3}\frac{\partial\mathcal{G}}{\partial v}\right]\\
    &=\left(\frac{L_z}{v}\right)^2\langle(\Delta v_\parallel)^2\rangle+\frac{r^2v^2\sin^2\theta-L_z^2}{2v^2}\langle(\Delta v_\perp)^2\rangle,\\
    \langle\Delta Q\rangle&=\Gamma'\left[\frac{2Q}{v}\frac{\partial\mathcal{H}}{\partial v}+\frac{Q}{v^2}\frac{\partial^2\mathcal{G}}{\partial v^2}\right.\\
    &\left.+\frac{1}{v^3}\left(r^2v^2(1+\cos^2\theta)-Q\right)\frac{\partial\mathcal{G}}{\partial v}\right]\\
    &=\frac{2Q}{v}\langle\Delta v_\parallel\rangle+\frac{Q}{v^2}\langle(\Delta v_\parallel)^2\rangle\\
    &+\frac{1}{2v^2}\left(r^2v^2(1+\cos^2\theta)-Q\right)\langle(\Delta v_\perp)^2\rangle,\\
    \langle (\Delta Q)^2\rangle&=\Gamma'\left\{\frac{4Q^2}{v^2}\frac{\partial^2\mathcal{G}}{\partial v^2}\right.\\
    &\left.+\left[\frac{4r^2}{v}\left(Q-L_z^2\cos^2\theta\right)-\frac{4Q^2}{v^3}\right]\frac{\partial\mathcal{G}}{\partial v}\right\}\\
    &=\frac{4Q^2}{v^2}\langle(\Delta v_\parallel)^2\rangle\\
    &+\left[2r^2\left(Q-L_z^2\cos^2\theta\right)-\frac{2Q^2}{v^2}\right]\langle(\Delta v_\perp)^2\rangle,
\end{aligned}
\label{eq: LDC}
\end{equation}
and the cross terms:
\begin{equation}
    \begin{aligned}
        \langle\Delta E\Delta L_z\rangle&=\Gamma'\left[L_z\frac{\partial^2\mathcal{G}}{\partial v^2}\right]=L_z\langle(\Delta v_\parallel)^2\rangle,\\
        \langle\Delta E\Delta Q\rangle&=\Gamma'\left[2Q\frac{\partial^2\mathcal{G}}{\partial v^2}\right]=2Q\langle(\Delta v_\parallel)^2\rangle,\\
        \langle\Delta L_z\Delta Q\rangle&=\Gamma'\left[\frac{2L_zQ}{v^2}\frac{\partial^2\mathcal{G}}{\partial v^2}\right.\\
        &\left.-\frac{2L_z}{v^3}\left(Q-r^2v^2\cos^2\theta\right)\frac{\partial\mathcal{G}}{\partial v}\right]\\
        &=\frac{2L_zQ}{v^2}\langle(\Delta v_\parallel)^2\rangle\\
        &-\frac{L_z}{v^2}\left(Q-r^2v^2\cos^2\theta\right)\langle(\Delta v_\perp)^2\rangle.
    \end{aligned}
    \label{eq: LDC cross terms}
\end{equation}

We first note that the energy diffusion coefficients, $\langle\Delta E\rangle$ and $\langle(\Delta E)^2\rangle$, obtained for a spherical and isotropic stellar field, are identical to those in the standard 2D case \citep{Merritt2013}. The diffusion coefficients associated with $L_z$, have also appeared in the literature \citep{Einsel&Spurzem1999,Fiestas+2006, Merritt2013}, although they have not been as widely used.

The new 3D diffusion coefficient, $\langle\Delta Q\rangle$, can be related to the 2D coefficients by considering the non-spinning limit of Kerr spacetime. In the limit $\chi_\bullet = 0$, the integrals of motion satisfy $Q+L_z^2=L^2$. 
This implies that $\langle\Delta Q\rangle+\langle\Delta(L_z^2)\rangle=\langle\Delta(L^2)\rangle$ where all quantities in angle brackets denote \textit{first} order diffusion coefficients. A similar relation holds for the second order terms. The relation between $\langle\Delta(L^2)\rangle$ and $\langle\Delta L\rangle$ can be computed analytically as
\begin{equation}
\langle\Delta(L^2)\rangle = 2L\langle\Delta L\rangle + \langle(\Delta L)^2\rangle,
\end{equation}
where the final term on the right-hand side is a second-order diffusion coefficient. An identical expression holds upon replacing $L$ with $L_z$. 
Therefore, the 3D diffusion formalism naturally reduces to the 2D case in the limit $\chi_\bullet=0$ using: 
\begin{equation}
\langle\Delta Q\rangle+2L_z\langle\Delta L_z\rangle+\langle(\Delta L_z)^2\rangle=2L\langle\Delta L\rangle+\langle(\Delta L)^2\rangle.
\end{equation}

The new local diffusion coefficients depend not only on $r$ but also on $\theta$. In addition, because we derived the coefficients in three dimensions, they also depend on the orbital inclination. In this paper, we use two complementary definitions of the inclination, $I$ and $\iota$, which are as follows:
\begin{equation}
\begin{aligned}
Q&=L_z^2\tan^2{I} + \chi_\bullet^2(1 - E^2)\sin^2{I}\\
Q&=L_z^2\tan^2{\iota}.
\end{aligned}
\end{equation}
The angle $I$ is simply related to the range of the angle $\theta$ through which an inclined orbit oscillates:
\begin{equation}
\theta_{\rm min} \le \theta \le \theta_{\rm max}\;,
\end{equation}
where $\theta_{\rm max} = \pi - \theta_{\rm min}$ and $I = \pi/2 - {\rm sgn}(L_z)\theta_{\rm min}$.  The angle $\iota$ is simply related to the integrals of motion $Q$ and $L_z$, and so is particularly easy to use. Across much of the bound-orbit parameter space, $\iota$ is approximately equal to $I$.  These angles are precisely equivalent in the Schwarzschild limit.
In Appendix \ref{appendix sec: GDC} (\cref{fig: LDC}), we show the dependence of the local diffusion coefficients on both $\theta$ and the inclination. Appendix \ref{appendix sec: alternative GDC} also derives several reparametrizations of the GDCs, e.g. in terms of various functions of inclination and $Q$; some of these will be more useful for different astrophysical problems, including the Monte Carlo algorithm we develop below.  

\subsection{Torus-Averaged Diffusion Coefficients}
\label{sec:torus}
To simplify the problem further, we assume that the timescale for changes in $f_b$ due to collisions is long compared with the orbital period. Under this assumption, the local diffusion coefficients can be replaced by their time averages over many orbits. Because the diffusion coefficients in  \cref{eq: LDC,eq: LDC cross terms} depend explicitly on $r$ and $\theta$, this averaging requires integration over both coordinates. 

Although the local diffusion is strongest near apocenter, the orbit also reaches closer to the MBH near pericenter, where relativistic effects become important. Therefore relativistic effects are relevant when performing the orbit averaging. We therefore model the motion of the test objects along Kerr geodesics \citep{Bardeen+1972}, rather than Keplerian orbits, when performing the orbit averaging. 

The full derivation of this orbit-averaging procedure is presented in Appendix \ref{appendix sec: orbit average}, and the resulting orbit-averaged value of a general diffusion coefficient $\langle\Delta V^i\rangle$ where $V^i\in\{E,L_z,Q\}$ is given by:
\begin{equation}
\begin{aligned}
    \langle\Delta V^i\rangle_{\rm o}&\equiv\frac{\int_{\theta_{\rm min}}^{\theta_{\rm max}}\int_{r_{\rm p}}^{r_{\rm a}} J\langle\Delta V^i\rangle drd\theta}{\int_{\theta_{\rm min}}^{\theta_{\rm max}}\int_{r_{\rm p}}^{r_{\rm a}}Jdrd\theta},\\
    J&\equiv\frac{\Sigma(r,\theta)^2}{r^2 \sqrt{R(r)}\sqrt{\Theta(\theta)}}
\end{aligned}
\end{equation}
where the functions $R(r)$ and $\Theta(\theta)$ are the radial and polar potentials of the Kerr metric \citep{Bardeen+1972}, and $\Sigma=r^2+\chi_\bullet^2\cos^2\theta$. Here, $r_{\rm p}$ and $r_{\rm a}$ are the pericenter and apocenter, and $\theta_{\rm min}$ and  $\theta_{\rm max}$ are the minimum and maximum values of $\theta$ along the orbit for a given inclination $I$, respectively. These turning points are the roots of the associated radial and polar effective potentials. 

\section{Gravitational Wave Losses}
\label{sec:GW} 

When the CO reaches pericenter it loses energy and angular momentum through GW emission. Detecting EMRIs with LISA will require highly accurate theoretical models of their waveforms in order to track the waves' phase evolution long enough to integrate them out of detector noise.  Developing such waveforms is an ongoing, active area of research; see \cite{Chua:2020stf, Katz:2021yft, Speri:2023jte, Chapman-Bird:2025xtd} for a status overview of the most ``analysis-ready'' waveform families. The difficulty arises from the need to calculate the orbital motion of the CO in the Kerr spacetime while accounting for radiation reaction. Because of the extreme mass ratio between the test object and the MBH, $m_\bullet/M_\bullet\ll 1$, the system can be modeled using BH perturbation theory with the Teukolsky equation \citep{Teukolsky1973}.  However, despite significant progress in understanding how to efficiently compute and use the data needed to generate these waveforms \citep{Hughes+2021}, the process remains computationally expensive, and the most accurate waveform models do not yet include fully generic configurations.  For past {\it LISA} data analysis studies, approximate waveform models have been developed that capture the main features of the signal \citep{Glampedakis+2002,Gair&Glampedakis2006,Babak+2007}.  We use these models for the astrophysical estimates that are considered in this work. 

In this paper, we use the approach of \cite{Gair&Glampedakis2006}, which assumes that the orbital parameters evolve adiabatically. Due to the extreme mass ratio, the orbital parameters change on a timescale much longer than the orbital period. Therefore, on short timescales the motion can be approximated as geodesic, neglecting the influence of gravitational back-reaction. Within this scheme, the orbital motion is described by an exact Kerr geodesic.  The orbit-averaged rates of change of the integrals of motion $(\Delta E_{\rm GW},\Delta L_{z,\rm GW},\Delta Q_{\rm GW})$ are computed with empirical formulae whose functional form resembles the post-Newtonian approximation, but which has been calibrated to reproduce strong-field point-particle results using the Teukolsky equation in certain limits (equatorial but eccentric orbits; and constant radius but inclined orbits).  These rates of change are then used to update geodesic parameters, generating a new geodesic. This procedure is iterated, thereby combining an exact description of the short-timescale dynamics with an approximate treatment of long-timescale radiation reaction. For this reason, this method is commonly referred to as a ``hybrid'' or ``kludge'' approach\footnote{This particular approximation is known as the ``numerical kludge,'' contrasting it with the ``analytic kludge'' model of \cite{Barack:2003fp}.}.

In Appendix \ref{appendix sec: GWs}, we show some examples of orbital evolution driven by GW emission, following \cite{Gair&Glampedakis2006}. Additionally, we compare the evolution of the semimajor axis and eccentricity due to GW emission, as given in \cite{Gair&Glampedakis2006}, with the leading-order analytical expression of the Peters' formula \citep{Peters1964}, which is widely used in many studies due to its simplicity (this leading-order formula uses Newtonian gravity to describe the short-timescale orbit kinematics, supplemented by the quadrupole formula to determine the long-timescale secular evolution).  An important point in the context of this work is that the orbital inclination does not evolve significantly under GW losses, a result that holds even deep into the strong field \citep{Hughes:1999bq, Hughes:2001jr, Drasco:2005kz}. Therefore, any change in inclination is expected to arise primarily from two-body relaxation.

\begin{figure}
    \includegraphics[width=85mm]{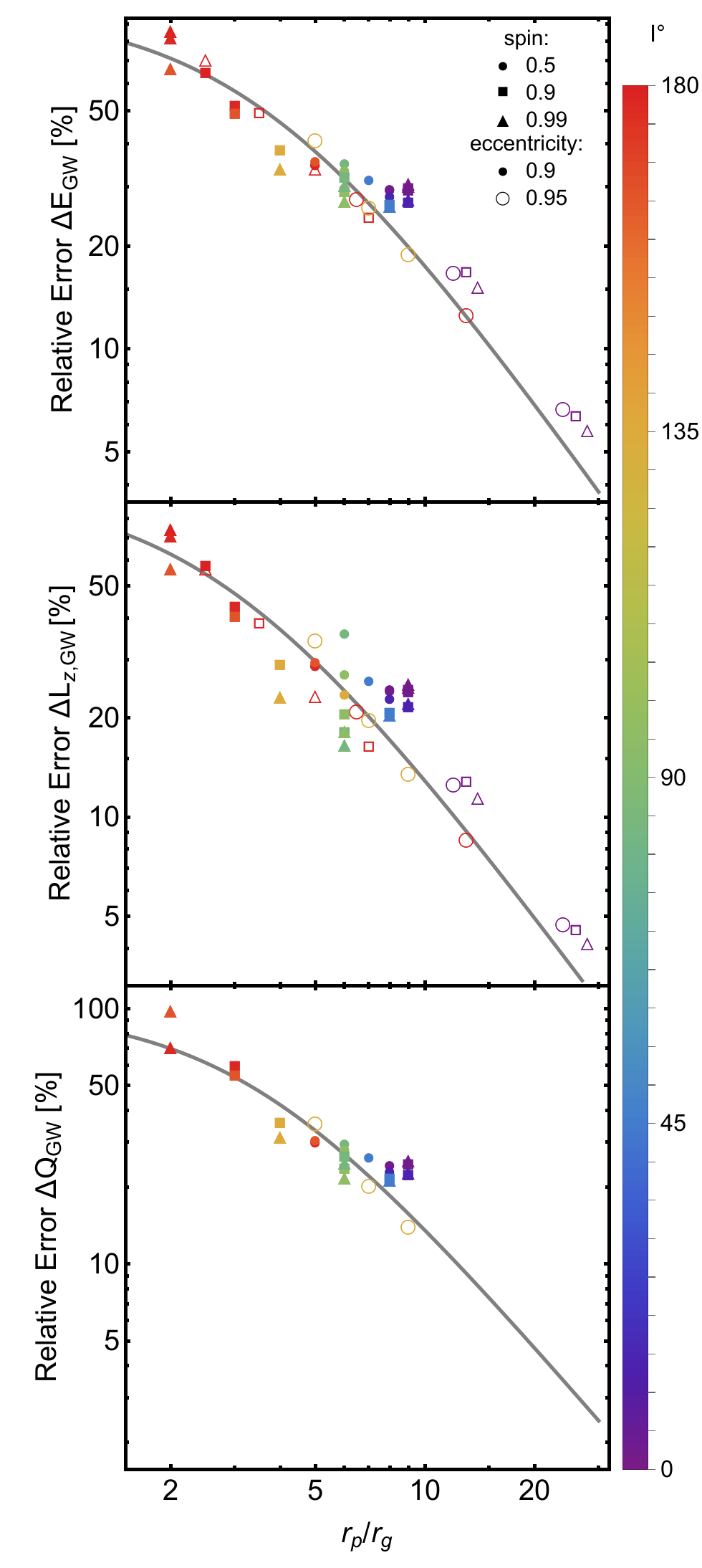}
\caption{The relative error in orbit-averaged changes to the energy (top), angular momentum in the z-direction (middle), and Carter constant (bottom) between the approximate GW approach \citep{Gair&Glampedakis2006} and the Teukolsky-based solution (\citealt{Teukolsky1973}, computed via the methods of \citealt{Hughes+2021}).  Results are shown as a function of pericenter distance, for different MBH spins, eccentricities and inclinations. The gray lines show power-law fits to the errors, which can be used to improve the accuracy of the \cite{Gair&Glampedakis2006} prescription.}
\label{fig: GG vs Teu}
\end{figure}

Since the method of \cite{Gair&Glampedakis2006} is based on a sequence of approximations, we test its accuracy by comparing it to results obtained using a more exact Teukolsky-based approach. Because of computational limitations, this approach has not yielded broad coverage of the relevant parameter space; however, it can be used without difficulty to calibrate the quality of our approximate formulas.  To this end, we use the code described in  \citet{Drasco&Hughes2006,Hughes+2021} that uses BH perturbation theory to compute GWs and the evolution of a geodesic's orbit integrals for a massive test body moving on bound geodesic orbits of Kerr black holes. The relative errors between these two approaches are shown in \cref{fig: GG vs Teu}. The comparisons are presented for the energy, the angular momentum in the z-direction, and the change in Carter constant, as functions of the pericenter radius, eccentricity, inclination, and MBH spin. For this comparison, we focus on high eccentricities and large MBH spins, where the approximation of \cite{Gair&Glampedakis2006} is expected to be least accurate.  \red{As shown in \cref{fig: GG vs Teu}, the relative error remains below $\sim 50\%$ as long as the pericenter radius $r_{\rm p}\gtrsim 3r_{\rm g}$.}
\blue{
As shown in \cref{fig: GG vs Teu}, the relative error in the flux remains modest ($\lesssim 10\%$) for distant pericenters ($r_{\rm p} \gtrsim 10 r_{\rm g}$), but increases for more relativistic orbits, reaching $\sim 50\%$ at $r_{\rm p} \approx 3 r_{\rm g}$ and up to a factor of $\approx 2$ ($\sim 100\%$ error) at the most extreme pericenters tested ($r_{\rm p} \approx 2 r_{\rm g}$).
}
While this level of accuracy would not be adequate for matched filtering or parameter estimation, it is more acceptable for astrophysical rate calculations.  Because the \citet{Gair&Glampedakis2006} kludge systematically {\it underestimates} GW losses, it implies that our results will be reasonably accurate for lower MBH spins, but may moderately underestimate EMRI rates for high spins where very small $r_{\rm p}/r_{\rm g}$ values are accessible.

\red{In \cref{fig: GG vs Teu}, the relative errors exhibit an approximately linear trend in log–log space as the pericenter radius decreases. Motivated by this behavior, we model the errors as a power law, $\epsilon_X(r_{\rm p})$, and compute the improved quantities as:
\begin{equation}
    \Delta X_{\rm improved}= \frac{\Delta X_{\rm G\&G}}{1-\epsilon_X(r_{\rm p})}
\end{equation}
where the best-fit error functions for the orbit-averaged GW changes of $X\in\{E,L_z,Q\}$ are given by:
\begin{align}
    \epsilon_E(r_{\rm p})   &= 1.34(r_{\rm p}/r_{\rm g})^{-0.802}, \\
    \epsilon_{L_z}(r_{\rm p}) &= 1.21(r_{\rm p}/r_{\rm g})^{-0.886}, \\
    \epsilon_{Q}(r_{\rm p})   &= 1.69(r_{\rm p}/r_{\rm g})^{-1.02}.
\end{align}
These fits provide a simple correction that can be used to improve the accuracy of the \cite{Gair&Glampedakis2006} prescription until broader strong-field parameter coverage for generic EMRI configurations becomes available.
}

\blue{
At large distances, the newly calculated changes are similar to those obtained using the prescription of \cite{Gair&Glampedakis2006}, but the two prescriptions increasingly diverge as $r_{\rm p}$ decreases. Motivated by this behavior, we fit the ratio between them and compute the improved quantities as:
\begin{equation}
    \Delta X_{\rm imp}= \Delta X_{\rm GG} F_X(r_{\rm p})
\end{equation}
where the best-fit correction functions for the orbit-averaged GW changes of $X\in\{E,L_z,Q\}$ are given by:
\begin{align}
    F_E(r_{\rm p})   &= 1+(7.11\pm1.18)\left(\frac{r_{\rm p}}{r_{\rm g}}\right)^{-1.53\pm 0.12}, \\
    F_{L_z}(r_{\rm p}) &= 1+(4.70\pm 0.70)\left(\frac{r_{\rm p}}{r_{\rm g}}\right)^{-1.50\pm 0.11}, \\
    F_{Q}(r_{\rm p})   &= 1+(7.34\pm 3.14)\left(\frac{r_{\rm p}}{r_{\rm g}}\right)^{-1.67\pm 0.29}.
\end{align}
These fits provide a simple correction that can, in principle, be used to improve the accuracy of the \cite{Gair&Glampedakis2006} prescription until broader strong-field parameter coverage for generic EMRI configurations becomes available. However, since the fits are based on a limited set of strong-field configurations, we do not use the improved values in our calculations and instead retain the original \cite{Gair&Glampedakis2006} prescription throughout this work; these fits are instead presented here solely for purposes of estimating the approximate error in our astrophysical calculations produced by our usage of the GW kludge.
}

\section{Monte Carlo Algorithm}
\label{sec:algo}
To study the effect of inclination diffusion on a test CO in the nuclear cluster, we perform orbit-averaged Monte Carlo (MC) simulations. The simulations include both the stochastic perturbations from two-body relaxation with other stars in the cluster and the deterministic losses due to GW emission. We neglect scalar resonant relaxation \citep{Rauch&Tremaine1996}, as previous studies have shown it to be relatively unimportant for EMRI rates \citep{Bar-Or&Alexander2016}.  We also neglect vector resonant relaxation \citep{Rauch&Tremaine1996}, \red{which has not previously been studied in this context} \blue{which has not previously been coupled into a live Monte Carlo loss-cone simulation}; its effect on EMRI rates in the Kerr spacetime may potentially be more important, and we discuss this point further in \S \ref{sec:discuss}.

To capture the inclination evolution of the test CO, we evolve the three integrals of motion: the orbital energy, $E$, the $z$-component of the angular momentum, $L_z$, and the Carter constant, $Q$. For the MC simulations, we adopt the prescription of \cite{ShapiroMarchant1978} with variable time steps. At each iteration, the updates to the integrals of motion are given by:
\begin{equation}
\begin{aligned}
    \delta E&=NP\Delta E_{\rm GW}+NP\langle\Delta E\rangle_{\rm o}+(NP\langle(\Delta E)^2\rangle_{\rm o})^{1/2}y_1,\\
    \delta L_z&=NP\Delta L_{z,\rm GW}+NP\langle\Delta L_z\rangle_{\rm o}+(NP\langle(\Delta L_z)^2\rangle_{\rm o})^{1/2}y_2,\\
    \delta Q&=NP\Delta Q_{\rm GW}+NP\langle\Delta Q\rangle_{\rm o}+(NP\langle(\Delta Q)^2\rangle_{\rm o})^{1/2}y_3,
\end{aligned}
\end{equation}
where each iteration corresponds to $N$ orbits with orbital period $P$. The timestep $NP$ is chosen such that \red{the change in each integral of motion during a given iteration remains a small fraction of its value from the previous step, while not being shorter than a single orbital period. In the case where $N=1$, we first evolve the integrals of motion due to GW emission near pericenter, and subsequently apply the effects of two-body scatterings near apocenter.}
\blue{the changes in the integrals of motion are no larger than a fraction $F_{\rm safe}=0.1$ of their current values. The resulting timestep is not, however, allowed to be shorter than one orbital period. When the timestep is shorter than one orbit, we instead set $N=1$ and apply the GW and scattering updates sequentially: we first advance the orbit by half a GW step, recalculate its orbital parameters, and then apply the scattering step.}
Here, $\Delta X_{\rm GW}$ denote the changes due to GW emission (\S \ref{sec:GW}), as in \cite{Gair&Glampedakis2006}, $\langle\Delta X\rangle_{\rm o}$ are the orbit-averaged diffusion coefficients from \S \ref{sec:GDC}, and $y_i$ are random variables drawn from a 3D multivariate Gaussian distribution with mean $\langle y_i\rangle=0$ and a variance-covariance matrix given by: 
\begin{equation}
    \begin{bmatrix}
    1&\frac{\langle\Delta E\Delta L_z\rangle_{\rm o}}{\sqrt{\langle(\Delta E)^2\rangle_{\rm o}\langle(\Delta L_z)^2\rangle_{\rm o}}}&\frac{\langle\Delta E\Delta Q\rangle_{\rm o}}{\sqrt{\langle(\Delta E)^2\rangle_{\rm o}\langle(\Delta Q)^2\rangle_{\rm o}}}\\
    \frac{\langle\Delta E\Delta L_z\rangle_{\rm o}}{\sqrt{\langle(\Delta E)^2\rangle_{\rm o}\langle(\Delta L_z)^2\rangle_{\rm o}}}&1&\frac{\langle\Delta L_z\Delta Q\rangle_{\rm o}}{\sqrt{\langle(\Delta E)^2\rangle_{\rm o}\langle(\Delta L_z)^2\rangle_{\rm o}}}\\
    \frac{\langle\Delta E\Delta Q\rangle_{\rm o}}{\sqrt{\langle(\Delta E)^2\rangle_{\rm o}\langle(\Delta Q)^2\rangle_{\rm o}}}&\frac{\langle\Delta L_z\Delta Q\rangle_{\rm o}}{\sqrt{\langle(\Delta L_z)^2\rangle_{\rm o}\langle(\Delta Q)^2\rangle_{\rm o}}}&1
    \end{bmatrix}.
\end{equation}

To ensure that the positive-definite quantity $Q$ remains physical during evolution, we instead evolve $\sqrt{Q}$. The corresponding diffusion coefficients, along with other alternative formulations, are presented in Appendix \ref{appendix sec: alternative GDC}.

In this work, we assume a single-mass stellar population with $m_\star=1M_\odot$, an isotropic velocity distribution, and a power-law mass density profile $\rho(r)=\rho_{\rm inf}(r/r_{\rm inf})^{-\gamma}$ where $r_{\rm inf}$ is the influence radius and $\rho_{\rm inf}$ is the mass density at $r_{\rm inf}$. 
The influence radius is defined as the radius enclosing twice the BH mass, and is empirically approximated as $r_{\rm inf}=GM_\bullet/\sigma^2$, where $\sigma$ is the stellar velocity dispersion in the galactic center. In this work, we use the $\sigma$ from the $M-\sigma$ relation, $\sigma=70(M_\bullet/1.53\times10^6M_\odot)^{1/4.24}$ km/s \citep{Gultekin+2009}.  
The gravitational potential is assumed to be $\Phi(r)=-GM_\bullet/r$, and using Eddington’s integral, the number density is
\begin{equation}
    f(E)=\frac{3-\gamma}{8}\sqrt{\frac{2}{\pi^5}}\frac{\Gamma(\gamma+1)}{\Gamma(\gamma-\frac{1}{2})}\frac{M_\bullet}{m_\star}\frac{\phi_0^{3/2}}{(GM_\bullet)^3}\left(\frac{|E|}{\phi_0}\right)^{\gamma-3/2} \label{eq:DF}
\end{equation}
where $\phi_0=GM_\bullet/r_{\rm inf}$. For the density slope, we use the \cite{Bahcall&Wolf1976} value $\gamma=7/4$ as a fiducial choice (but consider other cases later).  The use of a Keplerian potential limits us to consider the stellar distribution only at radii $r_{\rm g} \ll r \ll r_{\rm inf}$. At radii $r \sim r_{\rm g}$, Eq. \ref{eq:DF} will be inconsistent, but this matters little because here we are well inside the radii where stellar scatterings matter.  At radii $r \gtrsim r_{\rm inf}$, our neglect of the stellar potential will invalidate both Eq. \ref{eq:DF} as well as our torus-averaging procedure, so this limits the radial range over which we can consider EMRI formation.

The simulation is terminated under several stopping conditions. First, the integration is stopped if the test CO crosses the IBSO and undergoes a direct plunge. Second, it is terminated when the condition $t_{\rm GW}=10^{-3} t_{\rm rlx}$ \citep{Hopman&Alexander2005} is satisfied, indicating that the system has entered the GW–dominated regime and will evolve into an EMRI.
An additional termination criterion is applied when the CO reaches a large semimajor axis ($a > r_{\rm inf}/3$), in which case the orbit is classified as escaped and the simulation is stopped. Finally, because this work focuses on the diffusive regime, we exclude evolution into the high angular momentum (low eccentricity) region where dynamical friction dominates the orbital evolution. \red{Specifically, we terminate trajectories that enter the region where $\langle \Delta L \rangle < 0$, which we approximate by the condition $L/L_{\rm{c}} > 0.2 \sqrt{10 (m_\bullet/M_{\odot})^{-1} }$, where $L_{\rm{c}}$ is the circular angular momentum. This criterion is less restrictive than the low-eccentricity condition in \cite{Mancieri+2025}.}
\blue{In this region, the net drift $\langle \Delta L \rangle$ becomes negative due to dynamical friction drag, slowing diffusive angular momentum excitation toward the loss cone. We approximate the boundary where $\langle \Delta L \rangle \approx 0$ with the condition $L/L_{\rm{c}} > 0.2 (10 M_\odot/m_\bullet)^{1/2}$, where $L_{\rm c}$ is the circular angular momentum. For a canonical $m_\bullet=10M_\odot$ CO, this criterion corresponds to an eccentricity threshold of $e \lesssim 0.98$. Beyond this boundary, trajectories can evolve in energy space (through mass segregation) on timescales that are no longer short compared to angular momentum diffusion times, rendering the $S$-curve an ill-defined function. }

\section{Results}
\label{sec:results}







\begin{figure}
    \centering
    \includegraphics[width=85mm]{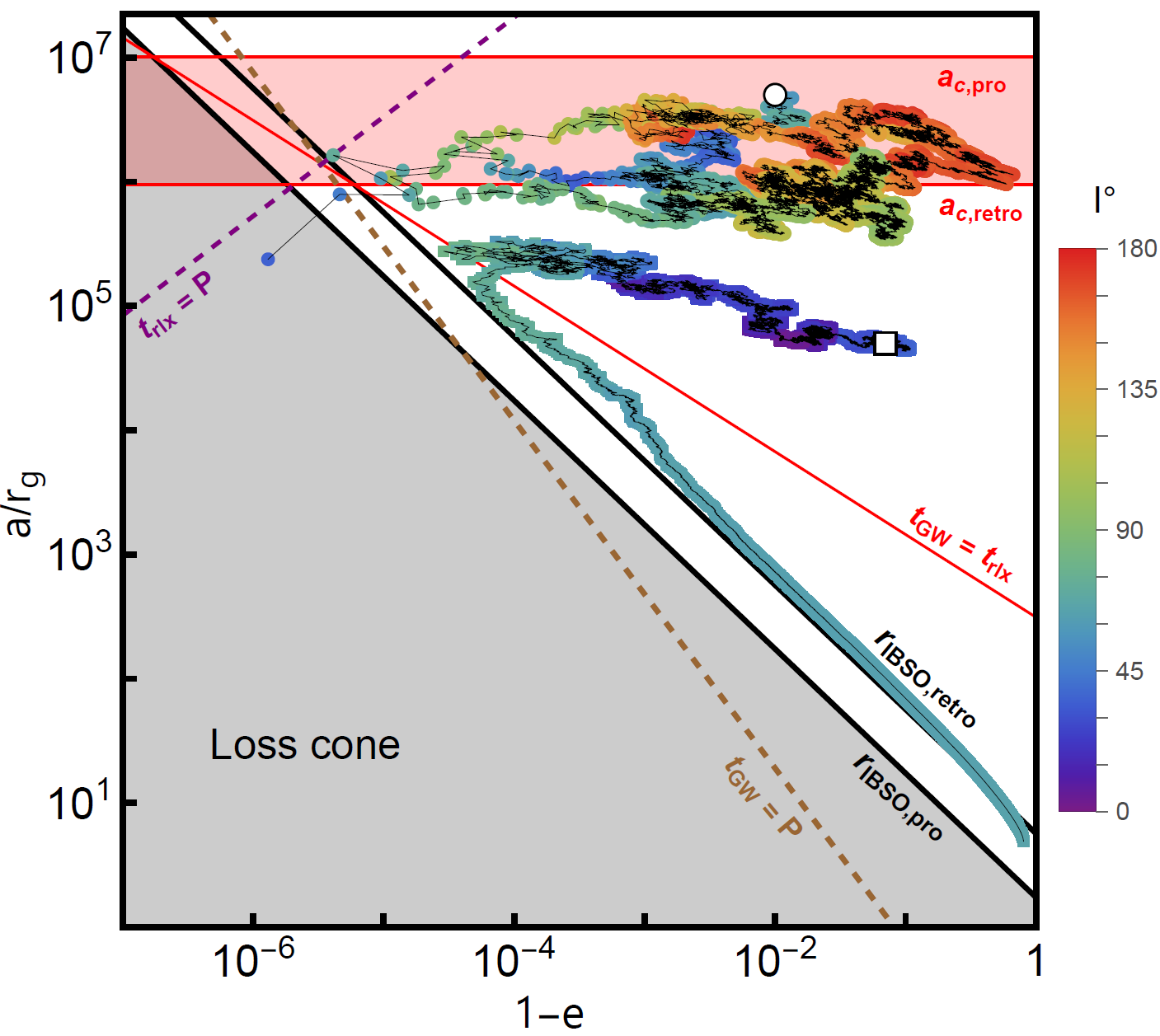}
    \includegraphics[width=85mm]{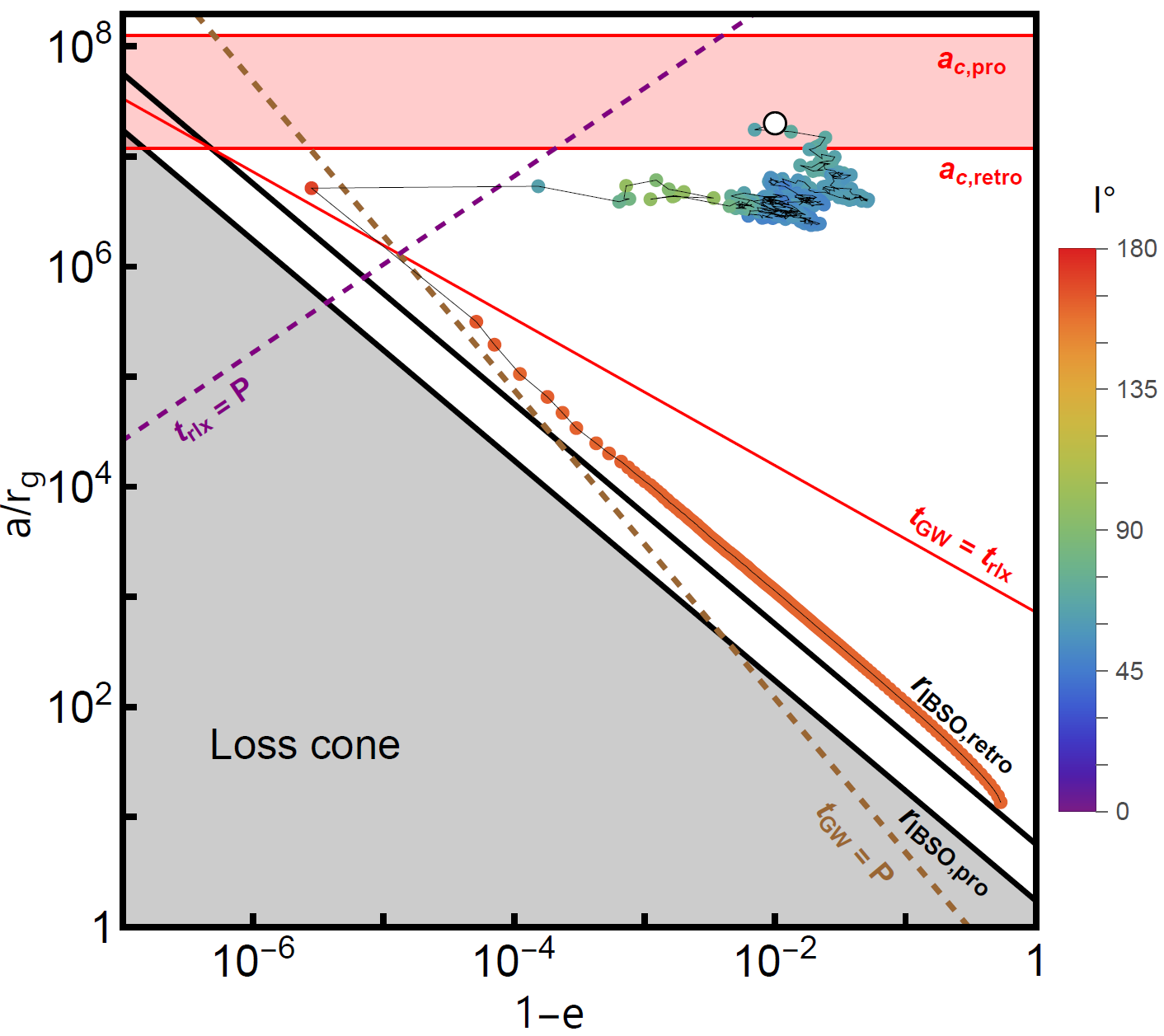}
    \caption{
    Example random walks from the 3D Monte Carlo simulation. The trajectories are shown as semimajor axis versus the eccentricity deficit $1-e$, with colors indicating the inclination ($0^\circ$ being prograde equatorial, $180^\circ$ retrograde equatorial). The maximum semimajor axis shown in each panel is the radius of influence. The spin of the MBH is $\chi_\bullet=0.9$, its mass is $M_\bullet=10^6M_\odot$ (top) or $M_\bullet=10^4M_\odot$ (bottom), and the mass of the CO is $m_\bullet=10M_\odot$.   
    The black solid lines are the IBSOs for equatorial prograde and retrograde orbits. The red curve shows where the GW timescale $t_{\rm GW}$ equals the angular momentum relaxation timescale $t_{\rm rlx}$, defining a range of critical semimajor axes, highlighted by the red region. In the top panel, we show two examples: a plunge orbit (starting at the circle), and an EMRI (starting at the square). In the bottom panel, we show a ``cliffhanger'' EMRI, which  takes a large step in semimajor axis when $t_{\rm GW}$ becomes comparable to the orbital period $P$ (brown dashed line). Another region that is more important for $M_\bullet = 10^4 M_\odot$ is where $t_{\rm rlx}$ becomes comparable to $P$ (purple dashed line). In this regime, the CO takes large steps in eccentricity and inclination. In all trajectories, inclination evolves significantly due to scatterings until GW emission becomes the dominant driver of orbital evolution.
    }
    \label{fig:EMRI vs Plunge}
\end{figure}

In this section, we present the results of our MC simulation investigating the effect of inclination diffusion. A previous study defined a critical semimajor axis separating EMRIs from plunges \citep{Hopman&Alexander2005}. However, since the IBSO depends on inclination when the MBH is spinning, adding the inclination diffusion transforms this critical semimajor axis into a range of semimajor axes. This range is shown as a shaded red region in \cref{fig:EMRI vs Plunge}, and it marks the zone where inclination diffusion has the strongest impact on EMRI dynamics.

\begin{figure}
    \centering
    \includegraphics[width=85mm]{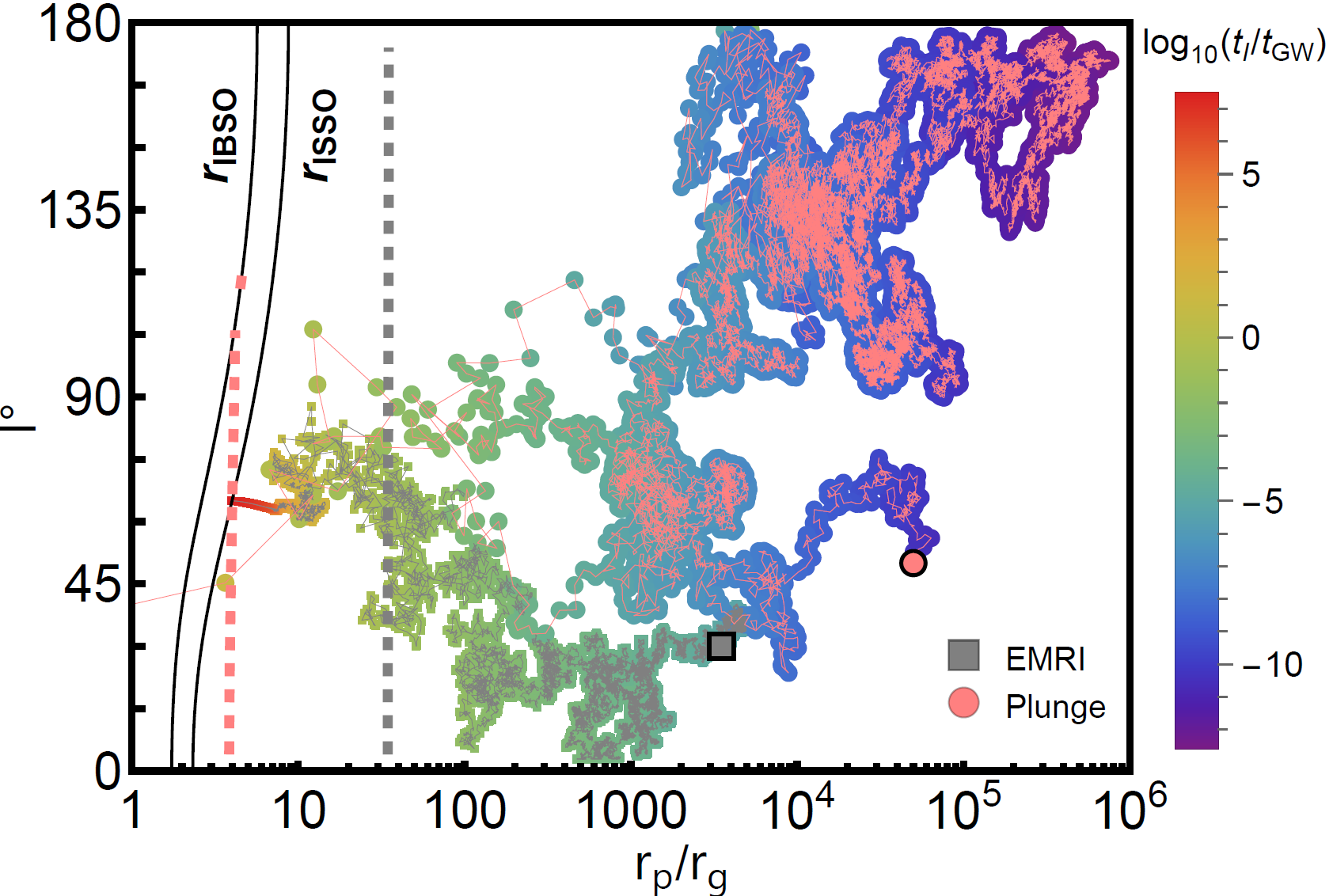}
    \caption{
    The same MC runs as in the top panel of \cref{fig:EMRI vs Plunge}, but now shown in the space of inclination and pericenter distance. The colors indicate the ratio of the inclination diffusion timescale to the GW inspiral timescale.     
    The black curves indicate the IBSO and ISSO. The two dashed lines show $t_{\rm GW}=t_\iota$ (Eqs. \ref{eq: GW timescale} and \ref{eq: inclination timescale}) for the initial semimajor axes of the plunge and EMRI. The figure demonstrates that the inclination can evolve significantly during the orbital evolution. For tightly bound orbits (e.g. classical EMRIs), inclination diffusion may become inefficient in the final stages of the random walk.
    }      
    \label{fig:EMRI vs Plunge i-rp}
\end{figure}

Two example runs from our MC code when the MBH mass is $M_\bullet=10^6 M_\odot$ are presented in \cref{fig:EMRI vs Plunge,fig:EMRI vs Plunge i-rp}. The simulations are initialized with semimajor axes, eccentricities, and inclinations of $(5\times 10^6r_{\rm g},0.99,50\degree)$ and $(5\times 10^4r_{\rm g},0.93,30\degree)$. In the first case, the orbit crosses the IBSO and results in a plunge, while in the second it enters the region where GW emission dominates over two-body scattering, leading to an EMRI. In both runs, the inclination diffuses due to two-body scatterings, which can flip the orbit from prograde to retrograde. Once the CO enters the GW-dominated regime, however, inclination diffusion halts.

If the central BH is an intermediate-mass black hole or the mass of the test CO is larger, a new class of EMRIs can arise: the so-called ``cliffhangers'' \citep{QunbarStone2024,Mancieri+2025}. In these cases, the orbit can undergo significant changes in its semimajor axis due to GW emission over a single orbital period. An example of such a cliffhanger is shown in the bottom panel of \cref{fig:EMRI vs Plunge}. Once the CO crosses the $t_{\rm GW} = P$ line, a cliffhanger can form.

To systematically investigate the effect of inclination diffusion on both the regular EMRIs and on the cliff hangers, we ran a large number of the MC simulations for a range of initial semimajor axes. In all runs, the initial eccentricity was fixed at $e_{\rm init}=0.999$, while the initial inclination was randomly sampled from \red{a uniform} \blue{an isotropic} distribution. Following \cite{Hopman&Alexander2005}, we estimate the EMRI-to-plunge ratio using the ``S-curves'', defined as
\begin{equation}
    S(a_{\rm init})=\frac{N_{\rm EMRI}}{N_{\rm EMRI}+N_{\rm plunge}}
    \label{eq: s-curves}
\end{equation}
where $N_{\rm EMRI}$ and $N_{\rm plunge}$ are the number of EMRIs and plunges, respectively, for a given initial semimajor axis $a_{\rm init}$. 

\begin{figure}
    \centering
     \includegraphics[width=85mm]{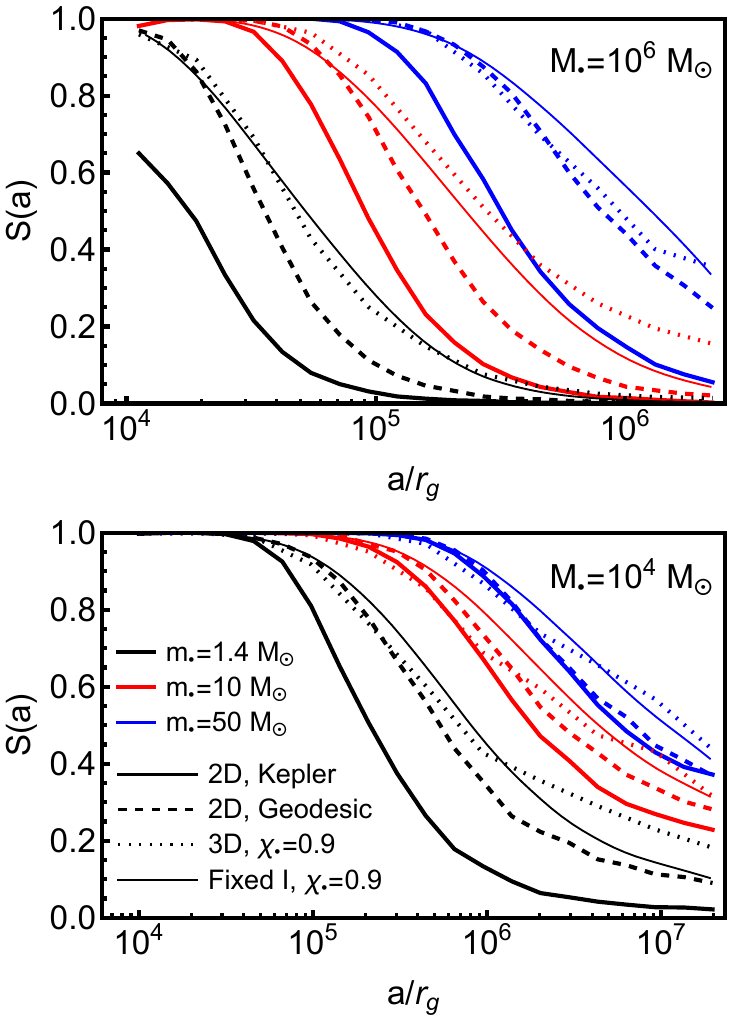}
    \caption{The EMRI fractions from all captures in \red{the} \blue{our} simulations as a function of the initial semimajor axis (i.e. the S-curves defined in Eq. \ref{eq: s-curves}). \red{We compare here the 2D and 3D cases, as well as various effects of GR.} Solid lines show \red{the} 2D \red{case} \blue{simulations} for Keplerian trajectories, \red{and} \blue{while} dashed lines show \red{the} 2D \red{case} \blue{simulations using} \red{for} (Schwarzschild) geodesics. In both cases, \blue{we use the} GW fluxes \red{follow} \blue{of} \citet{Peters1964}. Dotted lines show \red{the} 3D \red{case} \blue{simulations}, evolved on geodesics with GW fluxes from \citet{Gair&Glampedakis2006}, for a highly spinning MBH ($\chi_\bullet=0.9$). Thin solid lines show the equivalent fixed-inclination Kerr case investigated in \cite{AmaroSeoane+2013}, \blue{calculated here using \cref{eq: S-curve fixed incli}}. 
    The MBH mass is $M_\bullet=10^6M_\odot$ (top) or $M_\bullet=10^4M_\odot$ (bottom). The CO masses are: $m_\bullet=1.4M_\odot$ (black), $m_\bullet=10M_\odot$ (red), and $m_\bullet=50M_\odot$ (blue). Geodesic orbits yield more EMRIs than Keplerian orbits, and spinning MBHs yield more EMRIs than non-spinning MBHs. Inclination diffusion may act to increase or decrease the EMRI fraction compared to the fixed-inclination limit, depending on $M_\bullet$ and $m_\bullet$.  Differences between the underlying physical assumptions (i.e. Newtonian vs relativistic gravity, Schwarzschild vs Kerr, live vs fixed inclination) are of the least importance in the cliffhanger-dominated regime seen for $M_\bullet = 10^4 M_\odot$ and $m_\bullet = 50 M_\odot$.
    }
          
    \label{fig:scurves}
\end{figure}

In \cref{fig:scurves}, we show S-curves for both the 2D and 3D cases, considering two MBH masses, $M_\bullet\in\{10^4M_\odot,10^6M_\odot\}$, and three CO masses, $m_\bullet\in\{1.4M_\odot,10M_\odot,50M_\odot\}$. In the 2D case, only $E$ and $L$ are evolved, whereas in the 3D case $E$, $L_z$, and $\sqrt{Q}$ are evolved to account for inclination diffusion. To isolate the effect of evolving on Keplerian orbits versus geodesics, we present 2D results using both Keplerian trajectories and geodesics, with GW fluxes computed following  the formalism of \citet{Peters1964}. In the 3D case, the motion is along geodesics, and the GW fluxes are computed using \citet{Gair&Glampedakis2006}. For the 3D results, we adopt a highly spinning MBH with $\chi_\bullet=0.9$.

We also include a fixed-inclination case for comparison. The fixed-inclination S-curve is obtained by mapping the 2D geodesic results to a $\chi_\bullet = 0.9$ MBH using \red{\cref{eq: fixed I rate}} \blue{\cref{eq: fixed I ac} \citep{AmaroSeoane+2013}}, and then averaging over all inclinations:
\begin{equation}
S_{\rm fixed}(\chi_\bullet,a_{\rm init})=\frac{1}{2}\int^1_{-1}{\rm d}\cos{\iota} S_{\rm 2D,G}\big(a_{\rm eff}(\chi_\bullet,\iota)\big),
\label{eq: S-curve fixed incli}
\end{equation}
where $S_{\rm 2D,G}$ is the 2D S-curve computed with geodesics, and \red{$a_{\rm eff}=a_{\rm init}/\mathcal{W}^{\frac{20\gamma-45}{12-4\gamma}}(\chi_\bullet,\iota)$} \blue{$a_{\rm eff}=a_{\rm init}\mathcal{W}^{\frac{5}{6-4\gamma}}(\chi_\bullet,\iota)$} accounts for the dependence on spin and inclination.

In \cref{fig:scurves}, we first note a clear difference between the 2D cases evolved on Keplerian orbits and those evolved on geodesics. Naively, one might expect little to no difference between these approaches.  However, past work by \citet{Mancieri+2025} indicates that models of orbits which more accurately describe their strong-field behavior increase the number of EMRIs. To test this and to validate our 3D implementation, we computed an additional case, not shown in the figures (to avoid crowding), using the 3D code for a nearly non-spinning MBH ($\chi_\bullet=0.001$). We find only small differences between this case and the 2D geodesic results. This indicates that the discrepancy originates from evolving the orbit along a geodesic rather than a Keplerian trajectory, and not from the choice between 2D and 3D evolution or from the choice of GW flux prescriptions (\citealt{Peters1964} versus \citealt{Gair&Glampedakis2006}).
On geodesics, GW emission is stronger, leading to a higher EMRI fraction relative to plunges.

Next, we compare the non-spinning case with a rapidly spinning MBH, including the effects of inclination diffusion. From \cref{fig:scurves}, we see that increasing the MBH spin leads to a higher EMRI fraction. This enhancement is primarily driven by the shift in the IBSO between prograde and retrograde orbits, as shown by \citealt{AmaroSeoane+2013}. However, a major goal of this calculation is to determine the role of inclination diffusion. To isolate this effect, \cref{fig:scurves} also presents S-curves for spinning MBHs in which the inclination is held fixed in time. While the MBH spin alone increases the number of EMRIs, a clear difference emerges between cases with fixed inclination and those in which inclination evolves. This difference is most evident at large initial semimajor axes, corresponding to the range of critical semimajor axes shown in \cref{fig:EMRI vs Plunge}.

\begin{figure}
    \centering
     \includegraphics[width=85mm]{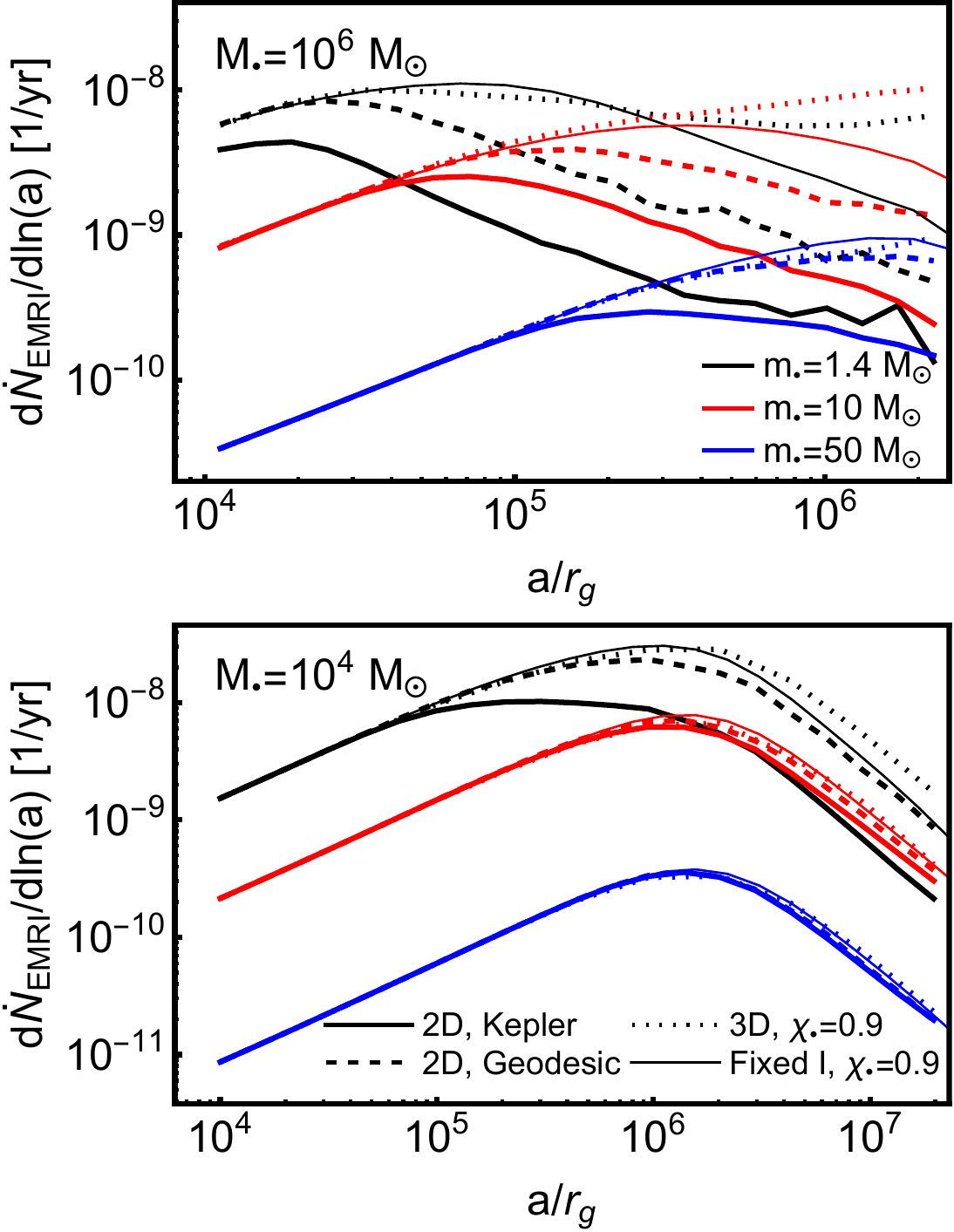}
    \caption{The differential EMRI rate per logarithmic semimajor axis (defined in Eq. \ref{eq:EMRI rates}) as a function of the initial semimajor axis. The line styles are the same as in \cref{fig:scurves}. The inclination diffusion cases deviate the most from the fixed-inclination cases at large semimajor axes. The MBH spin has little effect on the EMRI rates when cliffhanger EMRIs dominate the EMRI population.
    }
\label{fig:rates}

\end{figure}

\begin{figure}
    \centering
     \includegraphics[width=85mm]{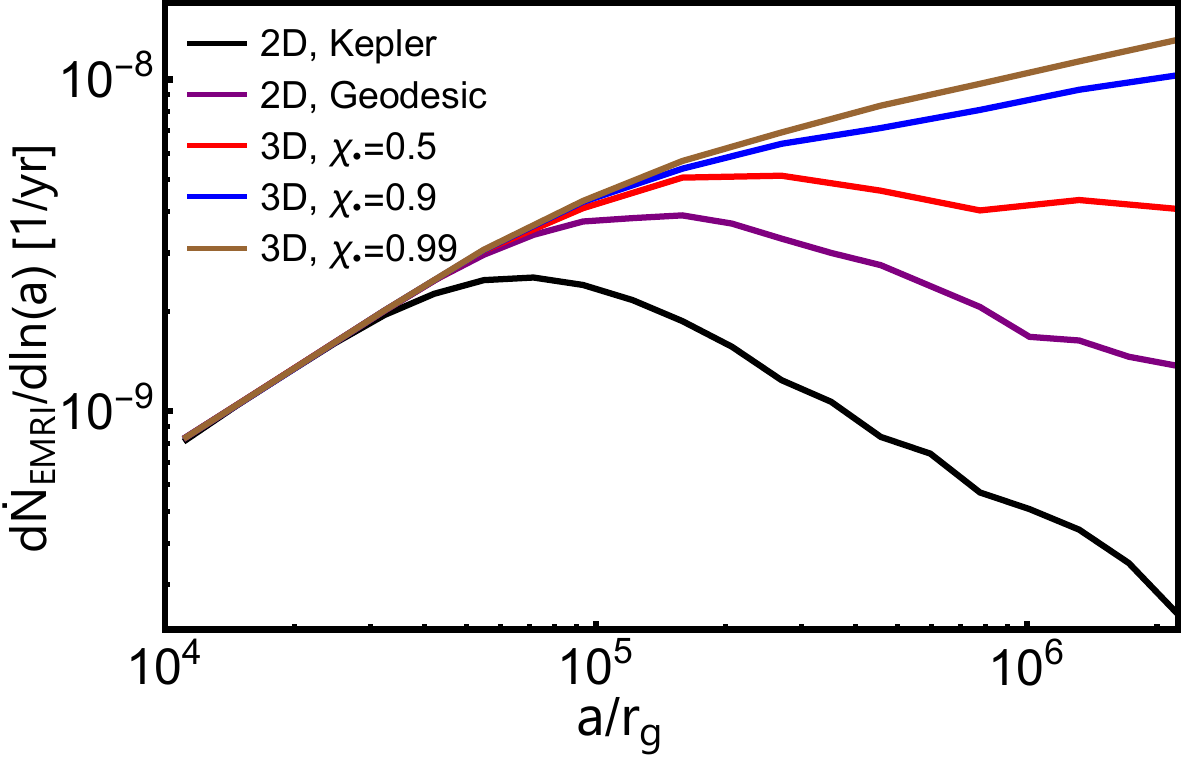}
    \caption{The differential EMRI rates, as in \cref{fig:rates}, for an MBH mass of $M_\bullet=10^6M_\odot$ and a CO mass of $m_\bullet=10M_\odot$. The rates are shown for 2D Keplerian orbits (black), 2D $\chi_\bullet = 0$ geodesics (purple), 3D orbits around a mildly spinning MBH with $\chi_\bullet=0.5$ (red), and 3D orbits around highly spinning MBHs with $\chi_\bullet=0.9$ (blue) and $\chi_\bullet=0.99$ (brown). As the MBH spin increases, so too does the EMRI rate, although this enhancement saturates at very high values of spin.
    }
\label{fig:rates M6 spins}
\end{figure}

\blue{These contrasts can also be seen in the EMRI rate profiles shown in \cref{fig:rates,fig:rates M6 spins}. The differential EMRI rate per logarithmic semimajor axis can be written as \citep{Merritt2013, Stone+2020}:
\begin{equation}
\begin{aligned}
   \frac{{\rm d}\dot{N}_{\rm EMRI}}{{\rm d}\ln{a}}&=\frac{{\rm d}N_\bullet(a)}{{\rm d}\ln{a}}\frac{S(a)}{P(a)}\times\\
   &\times
   \frac{q\mathcal{R}_{\rm IBSO}}{(\alpha-1)(1-\mathcal{R}_{\rm IBSO})+\ln{(1/\mathcal{R}_{\rm IBSO})}}
    \label{eq:EMRI rates}
\end{aligned}
\end{equation}
where ${\rm d}N_\bullet(a)/{\rm d}a=4\pi a^2 n_\bullet(a)$. Here,  $n_\bullet(a)=\eta_\bullet \rho_\bullet(a)/m_\bullet$ and $n_\star(a)=(1-\eta_\bullet) \rho_\star(a)/m_\star$ are the number densities of the COs and stars with masses $m_\bullet$ and $m_\star$, respectively, where $\rho_\bullet(a)\propto a^{-\gamma_\bullet}$, $\rho_\star(a)\propto a^{-\gamma_\star}$ are the corresponding mass densities. $\eta_\bullet$ is the number fraction of the COs in the cluster. For $m_\bullet=1.4M_\odot$ and $m_\bullet=10M_\odot$, we adopt $\eta_\bullet=10^{-2}$ \citep{HopmanAlexander2006Mass,RomSari2025}. For the more massive $m_\bullet=50M_\odot$, which are assumed to form through mergers, we adopt $\eta_\bullet=5\times 10^{-3}$. We define $\mathcal{R}_{\rm IBSO}=\left(L_{\rm IBSO}/L_c\right)^2$, where $L_c$ and $L_{\rm IBSO}$ are the angular momenta of circular orbits and of orbits at the IBSO, respectively. The dimensionless diffusivity of angular momentum relaxation is $q=\mathcal{D}P/\mathcal{R}_{\rm IBSO}$, and $\alpha=(q^2+q^4)^{1/4}$, where
\begin{equation}
\begin{aligned}
    \mathcal{D}(E)&=\frac{64\pi^2G^2\ln{\Lambda m^2_\star}(1-\eta_\bullet)}{3L^2_c(E)P(E)}\times\\
    &\times\int_0^{r_{\rm max}}{\rm d}r\frac{r^2}{\sqrt{2[E-\Phi(r)]}}(2\mathcal{I}_0+3\mathcal{J}_{1/2}-\mathcal{J}_{3/2}),\\
    \mathcal{I}_0(E)&=\int_E^0f_\star(E'){\rm d}E'\\
\mathcal{J}_n(E,r)&=\int_{\Phi(r)}^E\left[\frac{E'-\Phi(r)}{E-\Phi(r)}\right]^n f_\star(E'){\rm d}E'
\end{aligned}
\end{equation}
where $E$ is the orbital energy, $f_\star(E)$ is the stellar distribution function defined in \cref{eq:DF}, $\Phi(r)=-GM_\bullet/r$ is the gravitational potential, and $P(a)$ is the orbital period.}

In \cref{fig:rates}, we show the differential rates for the same cases as in the S-curves in \cref{fig:scurves}. As before, we find that the number of EMRIs is higher when the orbit is evolved along a geodesic rather than a Keplerian orbit, and increases further when the MBH is spinning. The rates show more clearly the difference between fixed and live inclination: the rates for these two cases are similar at low initial semimajor axes, but begin to diverge as the initial semimajor axis increases and approaches the range of critical semimajor axes. The rates also show that when cliffhangers dominate the EMRI population, there is little to no difference between the Schwarzschild and Kerr cases. We hypothesize that this is because cliffhangers can undergo large changes in semimajor axis over a single orbit, making the precise value of the IBSO less important for determining whether the CO becomes an EMRI. 
\blue{Geometrically, when the cliffhanger region in phase space is large, shifts in the loss-cone boundary driven by spin alter the overall area of the cliffhanger region only marginally, leading to negligible changes in the rate. 
}
\blue{We show the integrated EMRI rates in \cref{table: integrated rates}. These are obtained by integrating the differential rates over the semimajor axis range shown in \cref{fig:rates}.}

In \cref{fig:rates M6 spins}, we show $M_\bullet = 10^6 M_\odot$, $m_\bullet = 10 M_\odot$ EMRI rates for other values of the MBH spins.  In general EMRI rates increase with increasing MBH spin, though the effect saturates between $\chi_\bullet = 0.9$ and $\chi_\bullet = 0.99$.

\blue{
\begin{table}[h]
    \centering
    \begin{tabular}{|c||c|c|c|}
    \multicolumn{4}{c}{\blue{(a) $M_\bullet=10^6M_\odot$}}\\
    \hline
    & $m_\bullet=1.4M_\odot$ &$m_\bullet=10M_\odot$ &$m_\bullet=50M_\odot$ \\ \hline\hline
    2D Kepler& $7.4\times10^{-9}$ & $7.2\times10^{-9}$& $9.6\times10^{-10}$ \\ \hline
    2D Geodesics&$1.9\times10^{-8}$&$1.3\times10^{-8}$& $1.8\times10^{-9}$ \\ \hline
    3D $\chi_\bullet=0.5$&-&$1.9\times10^{-8}$& -\\ \hline
    3D $\chi_\bullet=0.9$&$4.0\times10^{-8}$&$2.8\times10^{-8}$& $1.9\times10^{-9}$\\ \hline
    3D $\chi_\bullet=0.99$&-&$3.1\times10^{-8}$&- \\ \hline
    \end{tabular}
    \begin{tabular}{|c||c|c|c|}
    \multicolumn{4}{c}{\blue{(b) $M_\bullet=10^4M_\odot$}}\\
    \hline
    & $m_\bullet=1.4M_\odot$ &$m_\bullet=10M_\odot$ &$m_\bullet=50M_\odot$ \\ \hline\hline
    2D Kepler& $4.2\times10^{-8}$ & $1.9\times10^{-8}$& $1.0\times10^{-9}$ \\ \hline
    2D Geodesics&$8.1\times10^{-8}$&$2.1\times10^{-8}$& $1.0\times10^{-9}$ \\ \hline
    3D $\chi_\bullet=0.9$&$1.0\times10^{-7}$&$2.1\times10^{-8}$& $1.0\times10^{-9}$\\ \hline
    \end{tabular}
    \caption{The integrated EMRI rates [1/yr] from \cref{fig:rates,fig:rates M6 spins}. }
    \label{table: integrated rates}
\end{table}
}

\begin{figure}
    \centering
     \includegraphics[width=85mm]{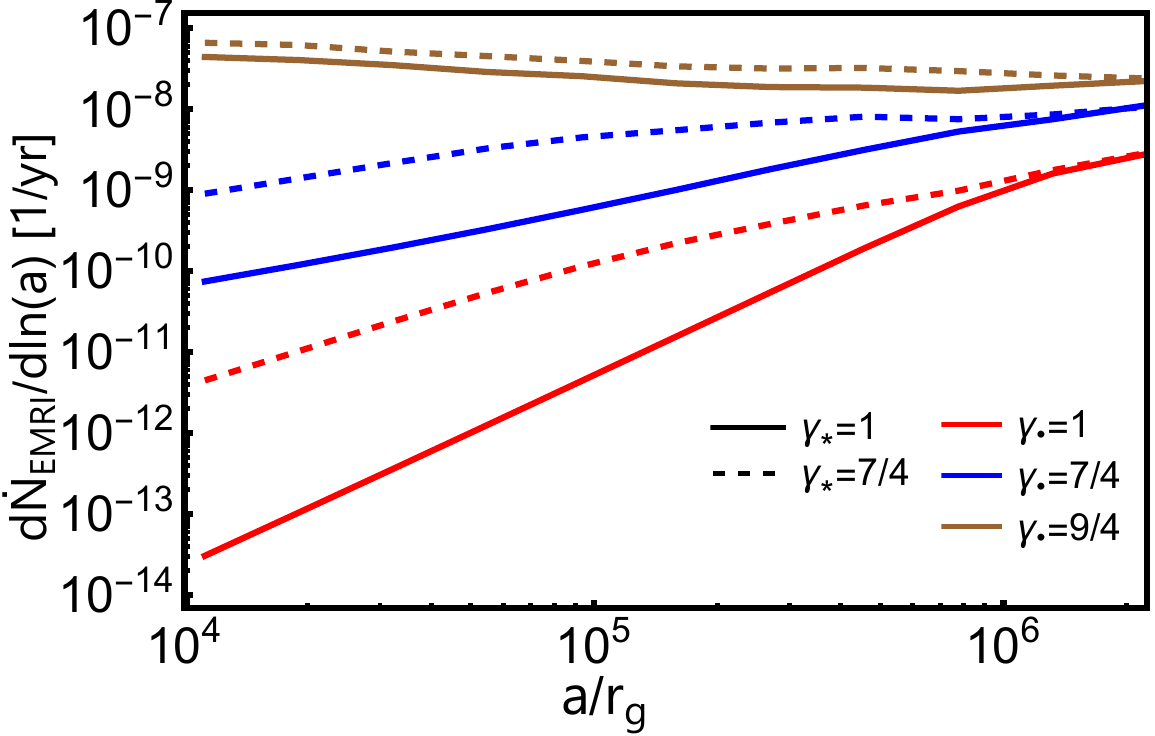}
    \caption{The differential EMRI rates, as in \cref{fig:rates}, for an MBH mass of $M_\bullet=10^6M_\odot$ and spin $\chi_{\bullet}=0.9$, and a CO mass of $m_\bullet=10M_\odot$. The rates are shown for stellar density power-law indices $\gamma_\star\in\{1,7/4\}$, and CO density power-law indices $\gamma_\bullet\in\{1,7/4,9/4\}$.
    As $\gamma_\bullet$ or $\gamma_\star$ increase, so too does the EMRI rate.
    }
\label{fig:rates M6 gammas}
\end{figure}

So far, we have considered scatterings from a single stellar population with a Bahcall–Wolf density profile ($\gamma = 7/4$). To explore how the EMRI rate depends on the slope $\gamma$, we also examine a two-component model in \cref{fig:rates M6 gammas}, consisting of $1M_\odot$ stars with slope $\gamma_\star$ and $10M_\odot$ COs with slope $\gamma_\bullet$, orbiting around a $M_\bullet=10^6 M_\odot$ MBH. We find that increasing $\gamma$ shifts the critical semimajor axis to smaller values, reducing the contribution of EMRIs formed at large semimajor axes. However, the overall EMRI rate increases with (either value of) $\gamma$, because the number of stars and COs at small semimajor axes grows as the density profile steepens.

\begin{figure}
    \centering
     \includegraphics[width=85mm]{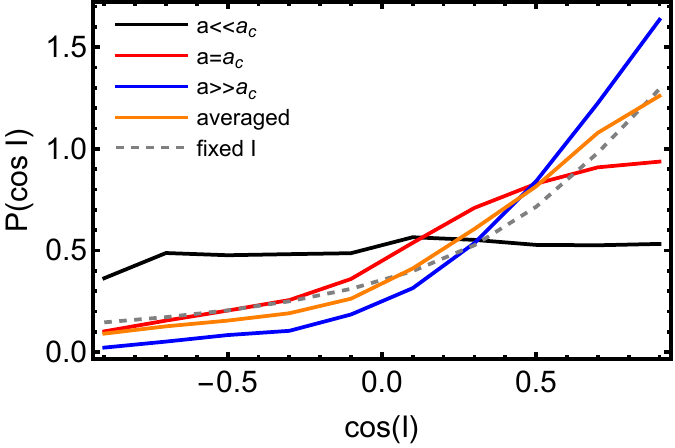}
    \caption{The inclination distribution of EMRIs for different initial semimajor axes. The MBH has spin $\chi_\bullet=0.9$ and mass $M_\bullet=10^6M_\odot$, and the CO has mass $m_\bullet=10M_\odot$. For small initial semimajor axes (black), the inclination distribution is isotropic because (i) nearly all orbits become EMRIs and (ii) the initial conditions are isotropic. As the initial semimajor axis increases, the EMRI inclination distribution becomes more prograde (red). At the largest semimajor axes (blue), the distribution is more strongly prograde than in the fixed-inclination case (dashed grey), although the final inclination distributions are similar when integrated over all semimajor axes (orange curve).
    }
\label{fig:incli dist}
\end{figure}

In \cref{fig:incli dist}, we show a fiducial inclination distribution of Kerr EMRIs ($M_\bullet = 10^6 M_\odot$, $\chi_\bullet = 0.9$, $m_\bullet = 10M_\odot$)  for different values of the initial semimajor axis. For very small semimajor axes ($a \ll a_{\rm c}$), the inclination distribution is isotropic. This occurs because (i) all orbits become EMRIs, so the IBSO does not play a role in selecting whether a system evolves into an EMRI, and (ii) our initial conditions are themselves isotropic. As the semimajor axis increases, the inclination distribution becomes more prograde-biased. For large semimajor axes ($a \gg a_{\rm c}$), inclination diffusion generates an even stronger prograde bias than in the fixed-inclination case; however, the $a$-integrated distribution remains similar to the fixed-inclination distribution.  

\begin{figure}
    \centering
     \includegraphics[width=85mm]{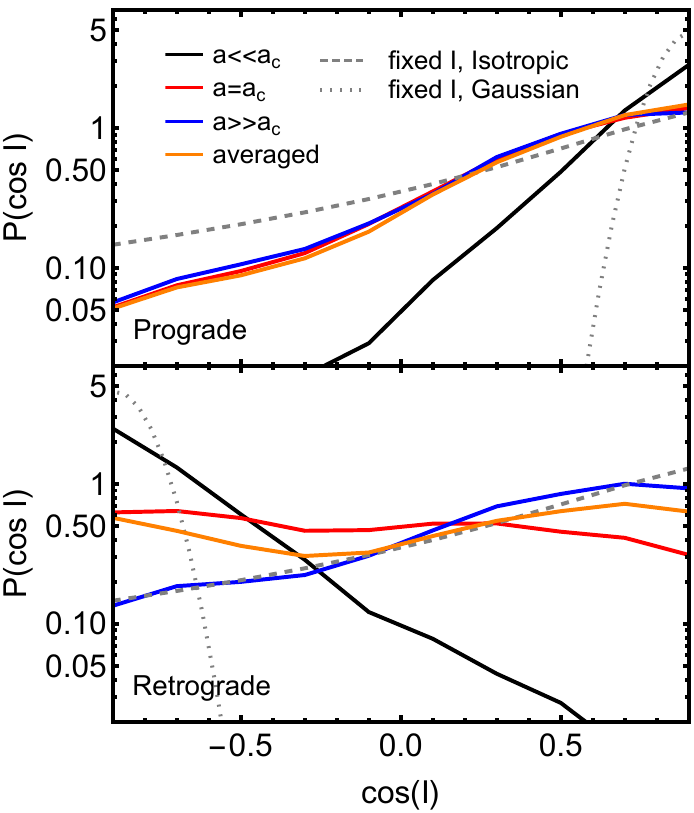}
    \caption{
    Inclination distribution of EMRIs for different initial semimajor axes, similar to \cref{fig:incli dist}, assuming initially prograde \blue{(top) and retrograde (bottom)} Gaussian distributions in $\cos I$ (\red{mean $0.9$} \blue{means $\pm0.9$}, widths $0.1$). At small initial semimajor axes, EMRIs remain strongly prograde\blue{/retrograde}. As the initial semimajor axis increases, inclination diffusion becomes efficient and \blue{almost} erases the initial anisotropy, yielding distributions similar to the $a\gg a_{\rm c}$ case in \cref{fig:incli dist}. For comparison, fixed-inclination models are shown: an isotropic case (less prograde than the diffusive results) and a \red{prograde} Gaussian case (remaining strongly prograde\blue{/retrograde}). This demonstrates that inclination diffusion efficiently washes out initial conditions at large semimajor axes.
    }
\label{fig:incli dist pro}
\end{figure}

During past AGN episodes, large numbers of COs may form {\it in situ} \citep{LevinBeloborodov03}; pre-existing COs may also become aligned with the gaseous disk through drag \citep{Syer+91, GenerozovPerets23}.  Even in vacuum, vector resonant relaxation may produce a ``dark disk'' of anisotropic, low-inclination heavy objects such as stellar mass black holes \citep{SzolgyenKocsis18}.  These effects may lead to \red{a preferentially prograde} \blue{an {\it initial} inclination distribution preferentially concentrated toward the equatorial plane} \footnote{More generally, the reasoning above suggests consideration of strongly anisotropic initial conditions for the CO population.  These anisotropies will be biased in the prograde equatorial direction if e.g. the most recent AGN episode succeeded at re-aligning the MBH spin vector with the orbital angular momentum of the AGN gas.  However, anisotropy out of the equatorial plane is possible if the AGN episode was short lived or involved little angular momentum exchange with the central MBH; likewise, dark disk anisotropies may in principle reflect the total angular momentum of the entire nuclear star cluster and not just the contribution of the most recent AGN episode.} \red{inclination distribution of {\it initial conditions}}. We therefore repeat this analysis assuming \red{initial conditions with a strong prograde bias} \blue{two strongly biased initial distributions, one favoring prograde orbits and the other favoring retrograde orbits} for the COs (though not for the field stars). As shown in \cref{fig:incli dist pro}, at small semimajor axes the EMRI population \red{ remains strongly prograde and closely reflects the initial conditions} \blue{retains a strong imprint of the initial inclination distribution}, although diffusion moderately broadens the distribution. As the semimajor axis increases, inclination diffusion becomes more efficient and works to erase memory of the initial anisotropy\red{. Consequently, at large semimajor axes the inclination distributions resemble those obtained for initially isotropic populations (\cref{fig:incli dist}), albeit with moderate differences.  Most notably a stronger suppression of retrograde orbits. This demonstrates that while an initially prograde configuration significantly affects EMRIs formed at small radii, its imprint is largely washed out at larger semimajor axes and in the average distribution. Additionally, fixed-inclination models, whether assuming isotropic or prograde Gaussian distributions, do not accurately capture this behavior.}
\blue{, causing the distributions to approach those obtained from initially isotropic populations (\cref{fig:incli dist}). Nevertheless, some differences remains: the prograde-biased model produces fewer retrograde EMRIs, while the retrograde-biased model retains an excess of retrograde EMRIs even at large semimajor axes. The mean inclination of the prograde-biased population is already close to that of the large semimajor axis isotropic population, whereas the retrograde-biased population retains a more noticeable retrograde component. Therefore, fixed-inclination models, whether assuming isotropic or prograde Gaussian distributions, do not accurately capture this behavior.}

\begin{figure}
    \centering
     \includegraphics[width=85mm]{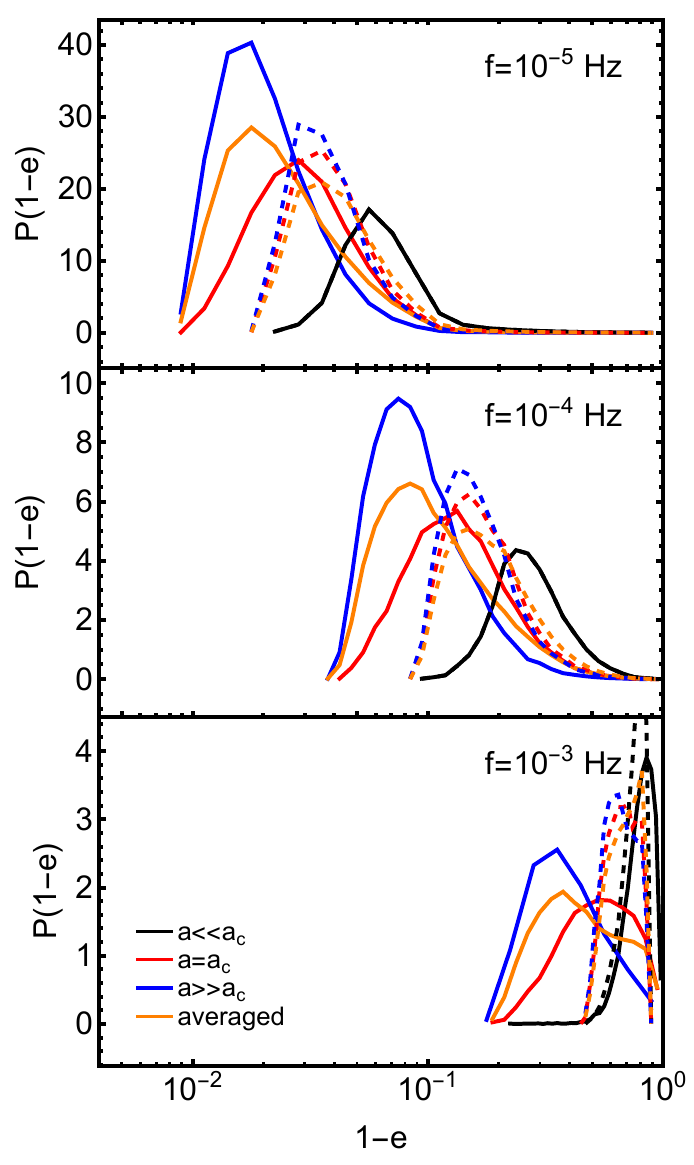}
    \caption{$1-e$ distribution of EMRIs for different initial semimajor axes. We show the distributions at different orbital frequencies relevant to {\it LISA}, $f\in\{10^{-5},10^{-4},10^{-3}\}$ Hz. The spin of the MBH is $\chi_\bullet=0.9$ (solid) or $\chi_\bullet=0$ (dashed). For small semimajor axes, the lines overlap. All other MBH and CO parameters are the same as in \cref{fig:incli dist}. The eccentricity distribution is largely determined by the distribution of initial semimajor axes, such that orbits with larger initial semimajor axes have higher eccentricities than those with smaller initial semimajor axes. The difference between the $\chi_\bullet = 0$ and $\chi_\bullet = 0.9$ cases is a consequence of this: as MBH spin allows EMRIs to preferentially originate from larger semimajor axes, they thus enter the {\it LISA} band with higher eccentricities.  As the EMRIs evolve through the {\it LISA} band, their orbits universally become more circular.
    }
\label{fig:ecce dist}
\end{figure}

In \cref{fig:ecce dist}, we show the eccentricity distribution for the same set of orbits presented in \cref{fig:incli dist}. As the systems evolve and enter deeper into the {\it LISA} band, their eccentricities decrease and the orbits become more circular. Here too, we find a dependence on the initial semimajor axis: orbits with larger initial semimajor axes retain higher eccentricities and are therefore less circular than those with smaller initial semimajor axes. Downstream of this observation, we also find a significant difference between the spinning and non-spinning cases. For large initial semimajor axes, the eccentricity distribution around a spinning MBH is shifted toward higher eccentricities compared to the non-spinning case, whereas for small semimajor axes there is no difference between the two distributions. 

\begin{figure}
    \centering
     \includegraphics[width=85mm]{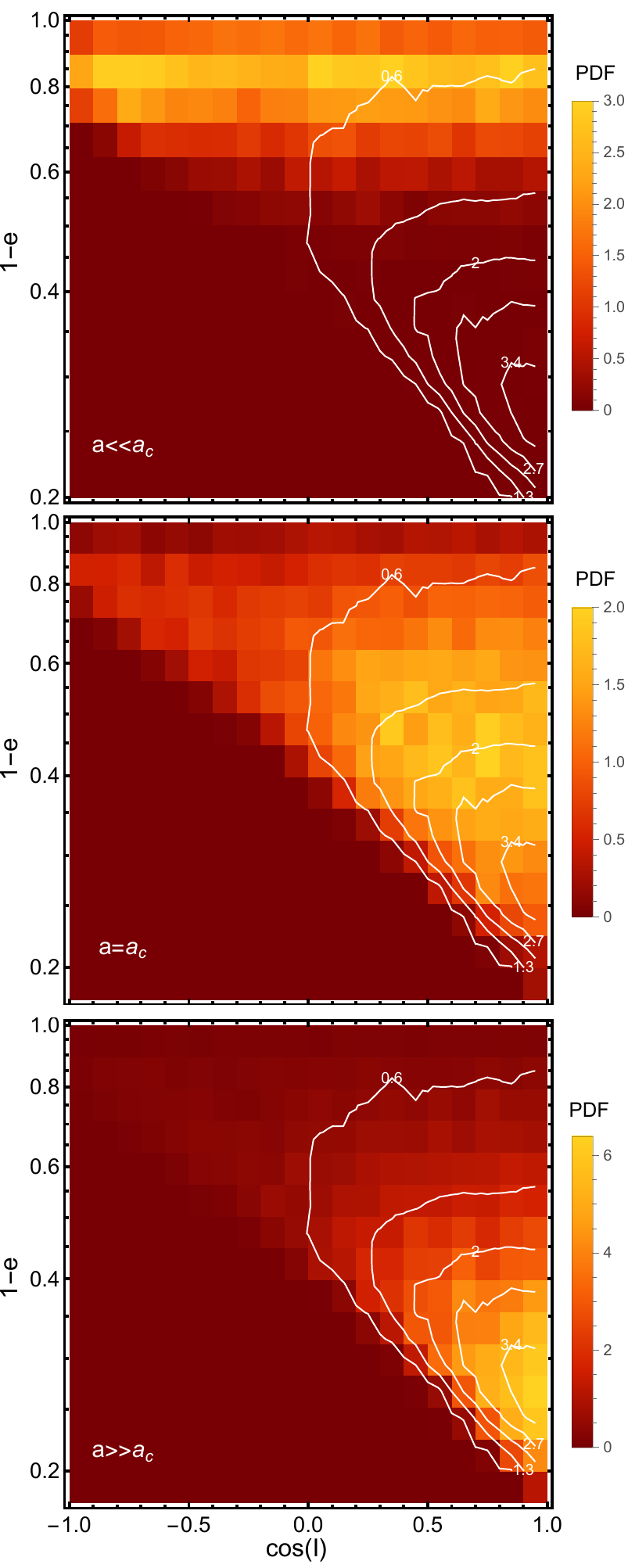}
    \caption{Bi-variate distributions of inclination $I$ and eccentricity deficity $1 - e$ for EMRIs at an orbital frequency of $f = 10^{-3}$ Hz. Colors are the (linear scale; see color bar) probability density function of the histograms. Each panel shows a different initial semimajor axis $a$, while the white contours indicate the $a$-integrated distributions. The MBH and CO parameters are the same as in \cref{fig:incli dist,fig:ecce dist}. 
    }
\label{fig:2D dist}
\end{figure}

We summarize our Monte Carlo runs by presenting bivariate distributions of inclination and eccentricity in \cref{fig:2D dist}.  This figure illustrates how the overall ``signature'' of loss cone EMRIs in the $\{e, I\}$ plane is ultimately a weighted average of sub-distributions that vary strongly with initial semimajor axis.  

\section{Discussion}
\label{sec:discuss}

\subsection{Geodesics versus (Post-)Newtonian EMRIs }

While the loss cone dynamics responsible for EMRI rates have been well studied in many previous works, most have operated with fundamentally Newtonian or post-Newtonian orbits \citep{HilsBender1995, SigurdssonRees1997, Hopman&Alexander2005, Bar-Or&Alexander2016, RavehPerets2021, QunbarStone2024, Mancieri+2025} rather than geodesics, and only a handful have considered the Kerr spacetime \citep{AmaroSeoane+2013,Vazquez-Aceves+2022, Sun+2026, ZhangAmaroSeoane2026}.  
A major ambiguity in non-geodesic loss cone calculations concerns the correct treatment of the IBCO (or IBSO).  The steep scaling of parabolic GW dissipation with pericenter means that $a_{\rm c}$, and thus the overall EMRI rate, carries a strong dependence on the innermost parabolic pericenter.  Within the (post-)Newtonian framework, multiple options exist for setting an IBCO, even for non-spinning BHs.  While one possible strategy could be to set $r_{\rm IBCO}=4 r_{\rm g}$, in alignment with the coordinate value of the geodesic IBCO in the Schwarzschild metric, this choice is somewhat arbitrary due to the gauge dependence of Schwarzschild coordinates \citep{ServinKesden17}.  A more common approach is to set $r_{\rm p} = 8 r_{\rm g}$ \citep{Hopman&Alexander2005}, so that the Newtonian angular momentum of a parabolic IBCO orbit equals the frame-invariant angular momentum of a parabolic IBCO geodesic in the Schwarzschild metric.  However, IBCOs also emerge in post-Newtonian gravity, and the recent work of \citet{Mancieri+2025} has used self-consistent (for $\chi_\bullet=0$) post-Newtonian IBCOs.

In \S\ref{sec:results}, we found that evolving orbits along geodesics, rather than Keplerian trajectories, leads to higher EMRI rates (as was previously suggested by the Newtonian to post-Newtonian progression in \citealt{Mancieri+2025}). This difference is not driven by the choice of 2D versus 3D evolution or by the GW flux prescription (i.e. \citealt{Peters1964} vs \citealt{Gair&Glampedakis2006}), but instead by the underlying orbital dynamics. In particular, geodesic evolution leads to stronger GW emission for fixed integrals of motion, which increases the chances of forming EMRIs over plunges. 
\blue{Mechanically, for fixed integrals of motion, geodesics exhibit closer pericenter passages than their Keplerian counterparts due to relativistic steepening of the effective potential. The resulting closer approach at pericenter produces stronger instantaneous GW emission, favoring circularization into an EMRI over direct plunge.}
The maximum EMRI rate enhancement we find by going from Keplerian orbits (with minimum $r_{\rm p} = 8 r_{\rm g}$) to Schwarschild geodesics is a factor $\approx 3$, though this can be much smaller, $\approx 1$, when $a \ll a_{\rm c}$ or when the Keplerian orbits are already deep in the cliffhanger regime.  In general, however, accurate EMRI rate estimates require relativistic geodesics.

\subsection{Inclination Diffusion}

Using an orbit-averaged Monte Carlo approach, we evolved Kerr geodesics in a 3D random walk across the space of $\{E, L_z, Q\}$.  To synthesize the results presented in greater detail in \S \ref{sec:results}, we summarize here a few broad results for EMRI rates in the Kerr spacetime:
\begin{itemize}
    \item When averaged over inclination, the S-curves for $\chi_\bullet > 0$ are always greater than the $\chi_\bullet = 0$ S-curves for $a \gtrsim a_{\rm c}$.  This spin enhancement becomes very large, roughly one order of magnitude, when $a \gg a_{\rm c}$ and the cliffhanger regime is unimportant (i.e. $t_{\rm GW} \gtrsim P$).  Conversely, the Schwarzschild and Kerr limits converge when $a \ll a_{\rm c}$.
    \item Deep in the cliffhanger regime, EMRI rates are extremely insensitive to MBH spin (see the $M_\bullet = 10^4 M_\odot$, $m_\bullet = 50 M_\odot$ case in Fig. \ref{fig:rates}).
    \item \red{EMRIs almost always arise from the empty loss cone regime, even when spin and/or cliffhanger effects allow them to originate from $a \gg a_{\rm c}$.}\blue{EMRIs always arise from the empty loss cone regime for $M_\bullet=10^6M_\odot$, whereas for $M_\bullet=10^4M_\odot$, they enter the full loss cone regime at large semimajor axes. }
    \item The inclination distribution of EMRIs in our fiducial simulations (with isotropic initial conditions) is a strong function of initial $a/a_{\rm c}$.  When $a/a_{\rm c} \ll 1$, the final EMRI ${\rm d}N/{\rm d}\cos (I) \approx 1/2$, consistent with the isotropy of our initial conditions.  When $a/a_{\rm c}\gg1$, the final EMRI ${\rm d}N/{\rm d}\cos (I)$ is highly anisotropic due to the strong $I$-dependence of the absorbing inner EMRI boundary condition.  Because most EMRIs are sourced from larger radii, the overall ${\rm d}N/{\rm d}\cos (I)$ is highly anisotropic for rapidly spinning MBHs.
    \item When looking at $a/a_{\rm c}\gg1$ EMRIs, the fiducial inclination anisotropy is somewhat stronger than the fixed-inclination case \citep{AmaroSeoane+2013}.  The greater anisotropy produced by inclination diffusion is evident for prograde EMRIs, which are modestly enhanced relative to the fixed inclination limit.  But it is even clearer for retrograde EMRIs, which are strongly suppressed due to the fixed inclination limit.  However, for our fiducial galactic nucleus model ($M_\bullet = 10^6 M_\odot$, $\gamma = 7/4$, $m_\bullet = 10 M_\odot$; see Fig. \ref{fig:incli dist}), the integral across all EMRI semimajor axes largely washes out the enhanced anisotropy from inclination diffusion, producing a final $a$-integrated ${\rm d}N/{\rm d}\cos(I)$ that is quite close to the prediction of \citet{AmaroSeoane+2013}.
    \item The role of inclination diffusion is most apparent for initially anisotropic populations. At $a/a_c \ll 1$, the inclination distribution largely preserves this initial prograde bias, with only modest broadening due to diffusion. In contrast, at $a/a_c \gg 1$, inclination diffusion becomes more efficient and rapidly erases the initial anisotropy, driving the system toward a distribution that is effectively indistinguishable from the isotropic case. Importantly, this behavior is not captured by fixed-inclination models, regardless of whether they assume isotropic or prograde initial conditions.    
    \item Eccentricity distributions are a strong function of spin.  Larger $\chi_\bullet$ values create a more radial eccentricity distribution at fixed pericenter frequency, {\it ceteris paribus}.
    \item \blue{The EMRI rate increases with increasing density profile slopes, $\gamma_\star$ and $\gamma_\bullet$, despite the critical semimajor axis decreasing, because steeper profiles concentrate more stars and COs at small semimajor axes.}
\end{itemize}


The comparison between our results and the fixed-inclination limit pioneered by \citet{AmaroSeoane+2013} is instructive, as it locates the regions of parameter space where $I$-diffusion plays a large role in determining CO capture outcomes, and those where it is less important.  
As mentioned above, the inclination distribution of EMRIs can be understood in terms of two asymptotic regimes.  When $a \ll a_{\rm c}$, essentially all CO captures will become EMRIs, and there is thus no way for the MBH spin to imprint itself into ${\rm d}N/{\rm d}I$.  Accordingly, we see that isotropic initial conditions yield isotropic outcomes for CO captures with $a \ll a_{\rm c}$.  This type of outcome distribution will also emerge when $a \gtrsim a_{\rm c}$, but EMRI dynamics are deep in the cliffhanger regime (which again makes spin effects unimportant).

In contrast, when $a \gg a_{\rm c}(\chi_\bullet=0)$ (i.e. the Schwarzschild critical radius), and cliffhangers are rare (i.e. $t_{\rm GW} \gg P$ always), then MBH spin effects are critical for EMRI production, and the strongly anisotropic inner boundary condition imprints itself into the inclination distribution.  What we observe in Fig. \ref{fig:incli dist} is that in this $a \gg a_{\rm c}(\chi_\bullet=0)$ regime, inclination diffusion leads to a {\it greater} anisotropy, namely a greater bias towards prograde EMRIs, than the fixed inclination approximation does.  This greater bias is moderate ($\approx 20\%$) for prograde orbits, but more dramatic for retrograde orbits.  Compared to the fixed inclination limit, we find that inclination diffusion reduces the highly retrograde EMRI fraction (again, for $a \gg a_{\rm c}(\chi_\bullet=0)$) by a factor $\approx 5$.

The greater prograde bias of live-inclination Monte Carlo runs is likely due to preferential elimination of moderate-inclination EMRIs by inclination diffusion, through a process that can be visualized using the $\{I, r_{\rm p} \}$ plane of Fig. \ref{fig:EMRI vs Plunge i-rp}.  For some COs that evolve stochastically from large semimajor axes down to very small $r_{\rm p}$, there will be a window of uncertainty as their pericenter approaches the $t_{\rm GW} = t_{\rm \iota}$ line denoting inclination freeze-out.  While $r_{\rm p}$ remains to the right of this line, inclination diffusion (i.e. vertical diffusion in Fig. \ref{fig:EMRI vs Plunge i-rp}) may convert the incipient EMRI into a plunge through an IBSO crossing.  The danger of this scenario is clearly higher for larger values of inclination (as these are closer to the IBSO in $\{I, r_{\rm p} \}$-space), and reduced as the CO approaches the prograde equatorial limit.

In contrast to our discussion so far, which reflects the outcomes of isotropic initial conditions for COs, one could imagine a strongly anisotropic set of initial conditions: for example, if the star cluster is rapidly rotating, then initial inclinations could be heavily biased towards prograde or retrograde distributions (depending on the misalignment between the MBH spin vector and the cluster net angular momentum vector).  An even stronger initial anisotropy might arise if most COs exist in a disk-like configuration, akin to the disk of O and B stars likely formed by Toomre instability in the Milky Way Center \citep{LevinBeloborodov03}, or alternatively, a dynamically formed dark disk produced by vector resonant relaxation \citep{KocsisTremaine2011}.  In these types of circumstances, for $a\ll a_c$ the inclination distribution is similar to the initial conditions. However, at larger semimajor axes, inclination diffusion rapidly isotropizes ${\rm d}N/{\rm d}\cos(I)$. 

Our calculations have been approximate in a number of ways that should be improved upon in future work.  First, our treatment of GW dissipation is simplistic, using the kludge approach of \citet{Gair&Glampedakis2006}.  By comparing to exact calculations of GW flux from the Teukolsky equation (see Fig. \ref{fig: GG vs Teu}), we have shown that this kludge systematically underestimates GW emission.  For pericenters $r_{\rm p} \gtrsim 10 r_{\rm g}$, this underestimate is modest enough to be unimportant for EMRI rates, producing an error $\lesssim 10\%$.  However, the discrepancy with the exact Teukolsky fluxes can rise to the factor of 2 level (i.e. $\approx 100\%$ error) for the most relativistic pericenters we tested ($r_{\rm p}\approx 2r_{\rm g}$).  This suggests that our $S$-curves may moderately underestimate the true EMRI fraction (and hence EMRI rate) from prograde equatorial orbits around the most rapidly spinning MBHs.  Work in progress \citep{Becker_inprep} aims to provide easy-to-use tools that provide accurate inspirals based on black hole perturbation theory using precomputed data describing orbit evolution over the parameter space.  It would be valuable to revisit these calculations when these tools are ready to be used in this analysis.

A second simplification in our calculation has been to use torus-averaged GDCs, in a geodesic extension of the standard (Newtonian) orbit-averaging procedure.  Past work has shown that orbit-averaging may underestimate EMRI rates compared to a more precise local calculation that follows a test CO along its orbit \citep{Mancieri+2025}.  While the comparison may not be precise (because of the post-Newtonian orbits used in \citealt{Mancieri+2025}), if this trend extends to geodesic diffusion, it suggests that torus-averaging may also contribute to a modest underestimation of the EMRI rate.

An additional limitation of torus-averaged GDCs is the implicit loss of phase-dependent interactions tied to the orbital orientation. For a given inclination, the instantaneous argument of pericenter dictates whether the compact object reaches pericenter near the equatorial plane or the poles. In non-spherical nuclear clusters, or in the Kerr metric, where spacetime curvature is axisymmetric, scattering rates and GW losses depending on this orientation. Torus-averaging assumes that relativistic precession cycles through all orbital orientations on a timescale much shorter than the relaxation time, smoothing these variations into a uniform background. 

Finally, we have employed relatively simple astrophysical models in our calculations: we neglect scalar and vector resonant relaxation \citep{HopmanAlexander2006, KocsisTremaine2011, Bar-Or&Alexander2016,Naoz&Haiman2023}, 
we compute isotropic diffusion coefficients (i.e. GDCs using an isotropic background of field stars), our assumption of spherical symmetry precludes collisionless eccentricity excitation \citep{VasilievMerritt2013, Penoyre+25}, we ignore loss cone shielding \citep{Teboul+24} from ejections in strong scatterings \citep{KaurPerets2024}, and we consider at most a 2-component population of stellar mass objects rather than a more realistic mass spectrum.  All of these astrophysical simplifications can in principle be relaxed in future work.

\subsection{Observational Implications}

There is currently little consensus on which channel for EMRI formation dominates overall event rates.  The loss cone channel we have explored in this work is the best studied of all proposed formation channels, but even here, EMRI rates are uncertain by at least three orders of magnitude \citep{Babak2017}. At least three other proposed channels may ultimately predominate in future EMRI samples.  First, the ``wet EMRIs'' produced via hydrodynamic CO inspiral in AGN accretion disks \citep{Levin2007} may reach large volumetric event rates if AGN efficiently form embedded COs through Toomre instability \citep{Pan+2021, DerdzinskiMayer2023}. Second, the Kozai-Lidov mechanism may likewise produce very high rates of EMRIs if MBH binaries are prevalent in most galactic nuclei \citep{Mazzolari+2022, Naoz+2022}.  Third, the deposition of COs onto tightly bound orbits via tidal separation of pre-existing binaries \citep{Hills88} may produce its own family of ``Hills'' EMRIs \citep{Miller+2005, RavehPerets2021}. There are even more EMRI formation channels that have been proposed in the past, such as EMRIs triggered by supernova kicks \citep{BortolasMapelli19}, but these have seen less detailed modeling.

Understanding the different contributions of different EMRI formation channels is a question of both astrophysical and physical interest.  Astrophysically, there are many long-standing open questions about the unknown populations of stellar mass COs in both quasi-spherical nuclear star clusters \citep{MiraldaEscudeGould00} and in AGN accretion flows \citep{Levin2007}; resolving either of these questions with an EMRI census of nuclear COs would have interesting implications for e.g. the formation of LIGO-band GW sources \citep{OLeary+09, GilbaumStone22}.  But such a census is, of course, only possible if we know which channel observed EMRIs come from.  Similar questions exist about stellar mass and MBH binary fractions in galactic nuclei, which could be probed by measurements of Hills and Kozai-Lidov EMRI rates, respectively.  

Physically, EMRIs have attracted great interest over the years as precision tests of general relativity and the Kerr hypothesis; for example, high-SNR EMRIs will measure MBH multipole moments and test the no-hair theorem to high precision \citep{Ryan1997}.  However, such tests will only be informative if we can be confident that the astrophysical environment is not altering the observed waveform from its vacuum template.  Such an assumption is likely reasonable for most EMRI channels, but might fail significantly for wet EMRIs, where hydro-gravitational torques can create significant phase shift in the waveform \citep{Yunes+11}.  This motivation further underscores the importance of isolating different formation channels, as {\it LISA} may fail to deliver on a primary science goal (tests of GR with EMRIs) if wet EMRIs cannot be confidently isolated from a future inspiral sample.

EMRIs from separate formation channels may be distinguished from each other, at least statistically, by ranges of orbital parameters.  Crudely speaking, the classical expectations for parameter distributions has been that loss cone EMRIs will have high eccentricities and quasi-isotropic inclinations; Hills EMRIs will have low eccentricities and quasi-isotropic inclinations; and wet EMRIs will have low eccentricities and low inclinations\footnote{Note, however, that if there is a significant fraction of retrograde inspirals through AGN disks, this wet EMRI subpopulation would have low inclinations but high eccentricities \citep{Secunda+21}}.  Our work here further substantiates past estimates in the fixed inclination limit demonstrating that the loss cone channel will have a substantial bias towards prograde equatorial orbits {\it if} rapidly spinning MBHs are abundant in the Universe \citep{AmaroSeoane+2013, Vazquez-Aceves+2022,Sun+2026, ZhangAmaroSeoane2026}.  Our examination of isotropization of strongly anisotropic initial conditions (Fig. \ref{fig:incli dist pro}) suggests that these results are relatively robust to the highly uncertain distribution of CO orbits in nuclear star clusters, and that the loss cone channel likely occupies a distinct region in $\{e, I\}$ parameter space (Fig. \ref{fig:2D dist}).

Even if all uncertainties on fractional contributions of different formation channels to the volumetric EMRI rate were resolved, this would not necessarily settle the issue of which channel will dominate the future {\it detection rate} of EMRIs.  {\it LISA} and {\it TianQin/Taiji} will not be equally capable of discovering all EMRIs.  Instead, their future EMRI samples will be shaped by a range of selection effects, many of which have been explored in other work.  Compared to a volume-complete EMRI sample, a strain-limited EMRI sample will be biased towards (i) larger secondary masses, (ii) a primary mass window in the $M_\bullet \sim 10^{5-6} M_\odot$ range, (iii) rapidly spinning MBHs, and (iv) prograde equatorial orbits.  This selection bias will further tilt observed samples of loss cone EMRIs towards an anisotropic $\cos I$ distribution (albeit, again, only if high MBH spins are prevalent).



Finally, we note that the geodesic diffusion coefficients derived here may be applicable to other relativistic loss cone problems, most notably tidal disruption events (TDEs) and quasi-periodic eruptions (QPEs).  For both of these electromagnetic transients, observable signatures may be sensitive to the inclination angle of the incoming star.  In QPEs sourced from star-disk collisions, the details of electromagnetic emission may depend on the star-disk misalignment angle \citep{Linial&Metzger2023}, in some cases strongly \citep{Tagawa&Haiman2023, Franchini+2023}.  In TDEs, Kerr inclination may imprint itself into global disk precession signatures \citep{Stone&Loeb2012, Franchini+2016}, circularization delays \citep{GuillochonRamirezRuiz15, Hayasaki+2016}, and even the binary question of whether or not a star disrupts at all \citep{Kesden12, Leloudas+16, Mummery24}.  While most past work on Kerr TDEs has assumed the fixed inclination limit, we note that \citet{SinghKesden24} additionally examined the limit of rapid isotropization.  In the final stages of the preparation of this manuscript, \citet{Xin2026} released a preprint deriving inclination diffusion coefficients equivalent to the Newtonian limit of those presented here. 
The work of \citet{Xin2026} to identify steady state solutions to the Fokker-Planck equation in the presence of inclination diffusion allows for a more realistic determination of Kerr TDE rates, and complements the Langevin-type analysis of this paper.

\section{Conclusions}
\label{sec:conc}

Estimates of EMRI rates remain highly uncertain, even for the most extensively studied formation channel, the loss cone \citep{Babak2017}. One source of this uncertainty is the spin of the MBH, as observational evidence indicates that a substantial fraction of MBHs in the Universe are rapidly rotating \citep{Reynolds2021}. The MBH spin can significantly influence the EMRI-to-plunge ratio, since the IBSO, which is the limiting radius for the last parabolic orbit, depends sensitively on the spin magnitude. For rapidly spinning MBHs, the IBSO decreases for prograde orbits, allowing COs to approach the MBH more closely without plunging and thereby increasing the EMRI formation rate. In contrast, for retrograde orbits, the IBSO increases, leading to an enhanced rate of direct plunges. Previous studies have demonstrated that highly spinning MBHs combined with low-inclination CO orbits can substantially increase the EMRI fraction \citep{AmaroSeoane+2013,Vazquez-Aceves+2022, Sun+2026, ZhangAmaroSeoane2026}; however, these estimates assumed a fixed orbital inclination throughout the orbital evolution.

However, two-body scatterings not only alter the energy and angular momentum of the CO but also modify its orbital inclination. Because the IBSO, and consequently the EMRI-to-plunge ratio, is sensitive to inclination in the presence of MBH spin, inclination evolution must be incorporated into accurate EMRI rate estimates. To this end, we developed the mathematical framework required to describe inclination diffusion driven by two-body encounters. Specifically, we generalize the conventional two-dimensional diffusion-coefficient formalism to three dimensions by evolving the three integrals of motion in the Kerr spacetime: the energy, $E$, the angular momentum component in the $z$-direction, $L_z$, and the Carter constant, $Q$ \citep{Bardeen+1972}, rather than the energy, $E$, and the total angular momentum, $L$, alone.

To investigate the impact of inclination diffusion in the presence of MBH spin, we incorporate the newly derived diffusion coefficients together with the \citet{Gair&Glampedakis2006} GW evolution  scheme into MC simulations. We find that the orbital inclination of the CO can evolve substantially over the course of the simulation, with initially prograde orbits becoming retrograde and vice versa, demonstrating the importance of inclination evolution. To quantify the effect on EMRI formation, we compute the S-curves, defined as the fraction of EMRIs over the total number of captures. From these, we derive the differential EMRI rate profile \citep{Hopman&Alexander2005}, as well as the inclination and eccentricity distributions. We find that transitioning from Keplerian trajectories to geodesic evolution, even in the 2D case, increases the EMRI fraction. Moreover, as expected, spinning MBHs produce more EMRIs than non-spinning ones, and inclination diffusion is most important for orbits with large initial semimajor axes. However, in the regime dominated by ``cliff-hanger'' EMRIs, the MBH spin has little impact on the overall EMRI rate, yielding results similar to the 2D geodesic case.  Overall, we find a fairly robust signature of the loss cone channel in the parameter space of eccentricity $e$ and inclination $I$: moderate-to-high values of $e$ along with a strong but not overwhelming prograde bias in $I$ (Fig. \ref{fig:2D dist}) will likely serve to distinguish loss cone EMRIs from those of other origins.

In this work, we improved the treatment of orbital inclination evolution driven by two-body scatterings; however, several additional dynamical processes operating in galactic nuclei were neglected. Previous studies have shown that scalar resonant relaxation is relatively unimportant for EMRI formation \citep{Bar-Or&Alexander2016}. In contrast, vector resonant relaxation \citep{Rauch&Tremaine1996}, which can significantly modify the inclination distribution of COs in the cluster, has not yet been explored in the context of EMRI rate estimates. We also neglect the gravitational potential of the surrounding stellar cluster and assume a spherical potential from the MBH. Deviations from spherical symmetry, such as axisymmetric or triaxial potentials, can induce anisotropic two-body scatterings that may alter our results \citep{VasilievMerritt2013,Kaur&Stone2025}. Furthermore, our simulations model scatterings with stars following a Bahcall–Wolf distribution. Using a more strongly mass-segregated distribution function could change the abundance of stellar-mass BHs and therefore the EMRI rates \citep{HopmanAlexander2006,Broggi+2024,Rom+2024, RomSari2025}. Moreover, COs on highly eccentric orbits can undergo ``zoom-whirl'' behavior, 
spending extended periods near pericenter before either plunging or returning to apocenter, and can therefore produce EMRI-like GW signals \citep{Schnittman2015,Naoz+2022}. Finally, the approximate gravitational-wave emission scheme employed in this work \citep{Gair&Glampedakis2006} can differ from Teukolsky-based calculations by tens of percent. Adopting a more accurate treatment \citep{Becker_inprep} may modify our results.

EMRI rates remain highly uncertain, and in this work we focus on one additional source of systematic uncertainty: the role of inclination diffusion in the presence of a spinning MBH. Our analysis was restricted to the loss-cone formation channel, however, the mathematical framework developed here for treating two-body scatterings in 3D is more general and may also be applicable to other EMRI formation channels, such as wet EMRIs in the presence of gas surrounding the MBH. Finally, inclination diffusion may also be relevant for other transients in galactic nuclei, such as tidal disruption events, for which loss-cone dynamics likewise play a central role \citep{Stone+2020, SinghKesden24, Xin2026}.

\section*{Acknowledgements}
The authors gratefully acknowledge helpful conversations with Cole Miller, Elena Rossi, Davide Mancieri, Luca Broggi, Pau Amaro Seoane, 
Adam Distler, Ismail Qunbar and Eytan Katzav.  SAH gratefully acknowledges helpful input from Diego Rivera Orona for major contributions to our program to compute gravitational waves from highly eccentric orbits. YA gratefully acknowledges the support of Azrieli fellowship. YA and NCS gratefully acknowledge support from the Binational Science Foundation (grant No. 2020397) and the Israel Science Foundation (Individual Research Grant No. 2414/23). Support for this research was provided by the University of Wisconsin–Madison, Office of the Vice Chancellor for Research and Graduate Education, with funding from the Wisconsin Alumni Research Foundation.  SAH's support for this work was provided by NSF Grant PHY–2409644.  Some of the computations which supported this analysis were done using the MIT {\tt engaging} cluster, which is supported by the MIT Office of Research Computing and Data, as well resources provided by {\tt subMIT} at MIT Physics.

\bibliography{main}{}
\bibliographystyle{aasjournal}

\appendix

\section{Inclination Relaxation Timescale}
\label{appendix: inclination timescale}
The inclination relaxation timescale is defined as the time required for an orbit to change its inclination by one radian:
\begin{equation}
    t_I\sim t_\iota=\frac{1}{\langle(\Delta\iota)^2\rangle}.
\end{equation}
where $\langle(\Delta\iota)^2\rangle$ is the second-order diffusion coefficient of the inclination. Under the assumptions in this paper, it is given by:
\begin{equation}
    \langle(\Delta\iota)^2\rangle=\frac{r^2(Q\sin^2\theta-L_z^2\cos^2\theta)}{2Q(Q+L_z^2)}\langle (\Delta v_\perp)^2 \rangle
\end{equation}
where $\cot\iota=L_z/\sqrt{Q}$, $Q$ is the Carter constant, and $L_z$ is the $z$-component of the angular momentum. In the far-field limit $L_z^2+Q=L^2$, where $L$ is the total angular momentum. 
The energy relaxation timescale is:
\begin{equation}
    t_E\sim \frac{v^2}{\langle (\Delta v_\parallel)^2\rangle}\sim\frac{2v^2}{\langle (\Delta v_\perp)^2\rangle}.
\end{equation}
Substituting, we obtain
\begin{equation}
   t_\iota\sim \frac{Q}{Q\sin^2\theta-L_z^2\cos^2\theta} \frac{L^2}{r^2v^2}t_E.
\end{equation}
Using $L_c \sim r v$, so that $L^2/L_c^2=1-e^2$, and the angular momentum relaxation timescale is $t_{\rm rlx}=t_E(1-e^2)$, the inclination timescale becomes:
\begin{equation}
   t_\iota\sim \frac{Q}{Q\sin^2\theta-L_z^2\cos^2\theta}t_{\rm rlx}=\frac{\sin^2{\iota}}{\sin^2{\iota}-\cos^2\theta}t_{\rm rlx}.
\end{equation}

To further simplify, we orbit-average the diffusion coefficient over $\theta$. As discussed in Appendix \ref{appendix sec: orbit average}, this requires integrating over $\theta$:
\begin{equation}
    \langle(\Delta \iota)^2\rangle_{\rm o,\theta}\equiv\frac{\int_{\theta_{\rm min}}^{\theta_{\rm max}}J_\theta\langle(\Delta \iota)^2\rangle {\rm d}\theta}{\int_{\theta_{\rm min}}^{\theta_{\rm max}}J_\theta {\rm d}\theta}
\end{equation}
where $J_\theta=\frac{1}{v_\theta}=\frac{Q}{\sqrt{1-\cot^2{\iota}\cot^2{\theta}}}$, neglecting the spin-dependent term. Using the change of variables $\sin{u}\equiv\frac{\cos{\theta}}{\sin{\iota}}$, we obtain:
\begin{equation}
    t_\iota\sim 2t_{\rm rlx}. 
\end{equation}

\section{General Geodesic Diffusion Coefficient Derivation}
\label{appendix sec: GDC}
To calculate the diffusion coefficients, we follow the approach of \cite{Rosenbluth+1957}. We start with the local Cartesian velocity coordinates: 
\begin{equation}
 q^1\equiv v=\sqrt{v_r^2+v_\theta^2+v_\phi^2},\, q^2\equiv v_\theta,\, q^3\equiv v_\phi.
\end{equation}
The corresponding metric tensor is: 
\begin{equation}
    a_{ij}=
    \begin{bmatrix}
    \left(\frac{v}{v_r}\right)^2 & \frac{-vv_\theta}{v_r^2} & \frac{-vv_\phi}{v_r^2} \\
    \frac{-vv_\theta}{v_r^2}   & \frac{v_r^2+v_\theta^2}{v_r^2} & \frac{v_\theta v_\phi}{v_r^2} \\
    \frac{-vv_\phi}{v_r^2}     & \frac{v_\theta v_\phi}{v_r^2} & \frac{v_r^2+v_\phi^2}{v_r^2}
\end{bmatrix}
\label{appendix eq: metric}
\end{equation}
with the determinant: $a\equiv\det(a_{ij})=v^2/v_r^2=v^2/(v^2-v_\theta^2-v_\phi^2)$.

The invariant form of the Fokker-Planck equation is:
\begin{equation}
\begin{aligned}
    \left(\frac{\partial f}{\partial t}\right)&=-\left(fT^i\right)_{;i}+\frac{1}{2}\left(fS^{ij}\right)_{;ij}\\
    &=-a^{1/2}\left[a^{1/2}f\left(T^i-\frac{1}{2}\Gamma_{jk}^iS^{jk}\right)\right]_{,i}+\frac{1}{2}a^{-1/2}\left[a^{1/2}fS^{ij}\right]_{,ij}
\end{aligned}
\end{equation}
where the semicolons denote covariant differentiation and commas indicate simple derivatives. $\Gamma^i_{jk}$ are the Christoffel symbols, and $T^i$ and $S^{ij}$ are the diffusion coefficients as tensorial objects: 
\begin{subequations}
\begin{align}
 T^{i}&=\langle\Delta v^i\rangle=\Gamma'a^{ij}\frac{\partial\mathcal{H}}{\partial v^j}\\
 S^{ij}&=\langle\Delta v^i \Delta v^j\rangle=\Gamma'\left[a^{ik}a^{jl}\left(\frac{\partial^2\mathcal{G}}{\partial v^k\partial v^l}- \Gamma_{kl}^m\frac{\partial \mathcal{G}}{\partial v^m}\right)\right]
\end{align}
\label{appendix eq: T S def}
\end{subequations}
where  $\Gamma'=4\pi G^2\ln{\Lambda}$ and
\begin{equation}
\begin{aligned}
    \mathcal{H}(\Vec{v})&=\sum_b m_b(m+m_b) h_b(\Vec{v}), \quad 
    \mathcal{G}(\Vec{v})=\sum_b m_b^2 g_b(\Vec{v}), \\ 
    h_b(\Vec{v})&=\int\frac{f_b(\Vec{v}_f)}{|\Vec{v}-\Vec{v}_f|} {\rm d}\Vec{v}_f, \quad   
    g_b(\Vec{v})=\int f_b(\Vec{v}_f)|\Vec{v}-\Vec{v}_f| {\rm d}\Vec{v}_f.
\end{aligned}
\end{equation}

The nonzero components of the Christoffel symbols:
\begin{equation}
\begin{aligned}
    \Gamma_{11}^1&=-\frac{v_\theta^2+v_\phi^2}{vv_r^2}, \quad    \Gamma_{12}^1=\Gamma_{21}^1=\frac{v_\theta}{v_r^2}, \quad
    \Gamma_{13}^1=\Gamma_{31}^1=\frac{v_\phi}{v_r^2},\\
    \Gamma_{22}^1&=-\frac{v^2-v_\phi^2}{vv_r^2}, \quad
    \Gamma_{23}^1=\Gamma_{32}^1=\frac{v_\theta v_\phi}{vv_r^2}, \quad
    \Gamma_{33}^1=-\frac{v^2-v_\theta^2}{vv_r^2}.
\end{aligned}
\label{appendix eq: Christoffel symbols}
\end{equation}

After substituting the metric and the Christoffel symbols from \cref{appendix eq: metric,appendix eq: Christoffel symbols} to \cref{appendix eq: T S def}, we get these expressions for $T^i$ and $S^{ij}$:
\begin{equation}
        T^1=\Gamma'\left(\frac{\partial\mathcal{H}}{\partial v}+\frac{v_\theta}{v}\frac{\partial\mathcal{H}}{\partial v_\theta}+\frac{v_\phi}{v}\frac{\partial\mathcal{H}}{\partial v_\phi}\right), \quad
        T^2=\Gamma'\left(\frac{v_\theta}{v}\frac{\partial\mathcal{H}}{\partial v}+\frac{\partial\mathcal{H}}{\partial v_\theta}\right),\quad
        T^3=\Gamma'\left(\frac{v_\phi}{v}\frac{\partial\mathcal{H}}{\partial v}+\frac{\partial\mathcal{H}}{\partial v_\phi}\right).
\end{equation}
\begin{equation}
    \begin{aligned}
        S^{11}&=\Gamma'\left[\frac{\partial^2\mathcal{G}}{\partial v^2}+\left(\frac{v_\theta}{v}\right)^2\frac{\partial^2\mathcal{G}}{\partial v_\theta^2}+\left(\frac{v_\phi}{v}\right)^2\frac{\partial^2\mathcal{G}}{\partial v_\phi^2}+2\frac{v_\theta}{v}\frac{\partial^2\mathcal{G}}{\partial v\partial v_\theta}+2\frac{v_\phi}{v}\frac{\partial^2\mathcal{G}}{\partial v\partial v_\phi}+2\frac{v_\theta v_\phi}{v^2}\frac{\partial^2\mathcal{G}}{\partial v_\theta\partial v_\phi}\right],\\
        S^{22}&=\Gamma'\left[\left(\frac{v_\theta}{v}\right)^2\frac{\partial^2\mathcal{G}}{\partial v^2}+\frac{\partial^2\mathcal{G}}{\partial v_\theta^2}+2\frac{v_\theta}{v}\frac{\partial^2\mathcal{G}}{\partial v\partial v_\theta}+\frac{v^2-v_\theta^2}{v^3}\frac{\partial\mathcal{G}}{\partial v}\right],\\
        S^{33}&=\Gamma'\left[\left(\frac{v_\phi}{v}\right)^2\frac{\partial^2\mathcal{G}}{\partial v^2}+\frac{\partial^2\mathcal{G}}{\partial v_\phi^2}+2\frac{v_\phi}{v}\frac{\partial^2\mathcal{G}}{\partial v\partial v_\phi}+\frac{v^2-v_\phi^2}{v^3}\frac{\partial\mathcal{G}}{\partial v}\right],\\
        S^{12}&=\Gamma'\left[\frac{v_\theta}{v}\frac{\partial^2\mathcal{G}}{\partial v^2}+\frac{v_\theta}{v}\frac{\partial^2\mathcal{G}}{\partial v_\theta^2}+\left(1+\left(\frac{v_\theta}{v}\right)^2\right)\frac{\partial^2\mathcal{G}}{\partial v\partial v_\theta}+\frac{v_\theta v_\phi}{v^2}\frac{\partial^2\mathcal{G}}{\partial v\partial v_\phi}+\frac{v_\phi}{v}\frac{\partial^2\mathcal{G}}{\partial v_\theta\partial v_\phi}\right],\\
        S^{13}&=\Gamma'\left[\frac{v_\phi}{v}\frac{\partial^2\mathcal{G}}{\partial v^2}+\frac{v_\phi}{v}\frac{\partial^2\mathcal{G}}{\partial v_\phi^2}+\frac{v_\theta v_\phi}{v^2}\frac{\partial^2\mathcal{G}}{\partial v\partial v_\theta}+\left(1+\left(\frac{v_\phi}{v}\right)^2\right)\frac{\partial^2\mathcal{G}}{\partial v\partial v_\phi}+\frac{v_\theta}{v}\frac{\partial^2\mathcal{G}}{\partial v_\theta\partial v_\phi}\right],\\
        S^{23}&=\Gamma'\left[\frac{v_\theta v_\phi}{v^2}\frac{\partial^2\mathcal{G}}{\partial v^2}+\frac{v_\phi}{v}\frac{\partial^2\mathcal{G}}{\partial v\partial v_\theta}+\frac{v_\theta}{v}\frac{\partial^2\mathcal{G}}{\partial v\partial v_\phi}+\frac{\partial^2\mathcal{G}}{\partial v_\theta\partial v_\phi}-\frac{v_\theta v_\phi}{v^3}\frac{\partial\mathcal{G}}{\partial v}\right].
    \end{aligned}
\end{equation}

To transform the diffusion coefficients from the coordinates $(v^1, v^2, v^3)$ to the coordinates $(V^1, V^2, V^3)$ we can use: 
\begin{equation}
\begin{aligned}
    \langle\Delta V^\lambda\rangle&=V^\lambda_{,i}T^i+\frac{1}{2}V^\lambda_{;ij}S^{ij}=\frac{\partial V^\lambda}{\partial v^i}T^i+\frac{1}{2}\left(\frac{\partial^2 V^\lambda}{\partial v^i\partial v^j}-\frac{\partial V^\lambda}{\partial v^k}\Gamma^k_{ij}\right)S^{ij},\\
    \langle\Delta V^\mu\Delta V^\nu\rangle&=V^\mu_{,i}V^\nu_{,j}S^{ij}=\frac{\partial V^\mu}{\partial v^i}\frac{\partial V^\nu}{\partial v^j}S^{ij}.
\end{aligned}
\label{appendix eq: transform LDC}
\end{equation}
where the $\mu, \nu, \lambda$ indices are associated with the new $(V)$ coordinate system and $i, j, k$ with the old $(v)$ system.

Now we can transform the diffusion coefficients from $(q^1,q^2,q^3)$ to $(E,L_z,Q)$ using \cref{appendix eq: transform LDC}:
\begin{equation}
\begin{aligned}
    \langle\Delta E\rangle&=\Gamma'\left[v\frac{\partial\mathcal{H}}{\partial v}+v_\theta\frac{\partial\mathcal{H}}{\partial v_\theta}+v_\phi\frac{\partial\mathcal{H}}{\partial v_\phi}+\frac{1}{2}\frac{\partial^2\mathcal{G}}{\partial v^2}+\frac{1}{2}\frac{\partial^2\mathcal{G}}{\partial v_\theta^2}+\frac{1}{2}\frac{\partial^2\mathcal{G}}{\partial v_\phi^2}+\frac{v_\theta}{v}\frac{\partial^2\mathcal{G}}{\partial v\partial v_\theta}+\frac{v_\phi}{v}\frac{\partial^2\mathcal{G}}{\partial v\partial v_\phi}+\frac{1}{v}\frac{\partial\mathcal{G}}{\partial v}\right],\\
    \langle(\Delta E)^2\rangle&=\Gamma'\left[v^2\frac{\partial^2\mathcal{G}}{\partial v^2}+v_\theta^2\frac{\partial^2\mathcal{G}}{\partial v_\theta^2}+v_\phi^2\frac{\partial^2\mathcal{G}}{\partial v_\phi^2}+2vv_\theta\frac{\partial^2\mathcal{G}}{\partial v\partial v_\theta}+2vv_\phi\frac{\partial^2\mathcal{G}}{\partial v\partial v_\phi}+2v_\theta v_\phi\frac{\partial^2\mathcal{G}}{\partial v_\theta \partial v_\phi}\right],\\
    \langle\Delta L_z\rangle&=\Gamma'\left[\frac{L_z}{v}\frac{\partial\mathcal{H}}{\partial v}+\frac{L_z}{v_\phi}\frac{\partial\mathcal{H}}{\partial v_\phi}\right],\\
    \langle(\Delta L_z)^2\rangle&=\Gamma'\left[\left(\frac{L_z}{v}\right)^2\frac{\partial^2\mathcal{G}}{\partial v^2}+\left(\frac{L_z}{v_\phi}\right)^2\frac{\partial^2\mathcal{G}}{\partial v_\phi^2}+2\frac{L_z^2}{vv_\phi}\frac{\partial^2\mathcal{G}}{\partial v\partial v_\phi}+\frac{L_z^2(v^2-v_\phi^2)}{v^3v_\phi ^2}\frac{\partial\mathcal{G}}{\partial v}\right],\\
    \langle\Delta Q\rangle&=\Gamma'\left\{\left(-2\frac{vE}{c^4}\cos^2\theta a^2+2r^2\frac{v_\theta^2}{v}+2r^2\cos^2\theta\frac{v_\phi^2}{v}\right)\frac{\partial\mathcal{H}}{\partial v}+\left(-2\frac{vE}{c^4}\cos^2\theta a^2\frac{v_\theta}{v}+2r^2v_\theta\right)\frac{\partial\mathcal{H}}{\partial v_\theta}\right.\\
    &\left.+\left(-2\frac{vE}{c^4}\cos^2\theta a^2\frac{v_\phi}{v}+2r^2\cos^2\theta v_\phi\right)\frac{\partial\mathcal{H}}{\partial v_\phi}+\left[r^2\left(\frac{v_\theta}{v}\right)^2+\cos^2\theta\left(-a^2(E+v^2)+r^2\left(\frac{v_\phi}{v}\right)^2\right)\right]\frac{\partial^2\mathcal{G}}{\partial v^2}\right.\\
    &\left.+\left(r^2-a^2\cos^2\theta\frac{E+v_\theta^2}{c^4}\right)\frac{\partial^2\mathcal{G}}{\partial v_\theta^2}  +\left(r^2\cos^2\theta-a^2\cos^2\theta\frac{E+v_\phi^2}{c^4}\right)\frac{\partial^2\mathcal{G}}{\partial v_\phi^2}    +2\frac{v_\theta}{v}\left(r^2-a^2\frac{E+v^2}{c^4}\cos^2\theta\right)\frac{\partial^2\mathcal{G}}{\partial v\partial v_\theta}\right.\\
    &\left.+2\frac{v_\phi}{v}\cos^2\theta\left(r^2-a^2\frac{E+v^2}{c^4}\right)\frac{\partial^2\mathcal{G}}{\partial v\partial v_\phi} -2a^2\frac{v_\theta v_\phi}{c^4}\cos^2\theta\frac{\partial^2\mathcal{G}}{\partial v_\theta\partial v_\phi}\right.\\
    &\left.+\frac{1}{v^3}\left(r^2(v^2-v_\theta^2)+\cos^2\theta(-2a^2\frac{Ev^2}{c^4}+r^2(v^2-v_\phi^2))\right)\frac{\partial\mathcal{G}}{\partial v}\right\},\\
    \langle (\Delta Q)^2\rangle&=\Gamma'\left\{\frac{1}{v^2}\left[-a^2\frac{Ev^2}{c^4}+r^2(2v_\theta^2+v_\phi^2)+\cos(2\theta)(-a^2\frac{Ev^2}{c^4}+r^2v_\phi^2)\right]^2\frac{\partial^2\mathcal{G}}{\partial v^2}\right.\\
    &\left.+v_\theta^2(2a^2\frac{E}{c^4}\cos^2\theta-2r^2)^2\frac{\partial^2\mathcal{G}}{\partial v_\theta^2}+4v_\phi^2\cos^4\theta(r^2-a^2\frac{E}{c^4})^2\frac{\partial^2\mathcal{G}}{\partial v_\phi^2}\right.\\
    &\left.+\frac{8}{v}\left[r^4v_\theta^3+\cos^2\theta(-a^2\frac{E}{c^4}r^2v^2v_\theta-a^2\frac{E}{c^4}r^2v_\theta^3+r^4v_\theta v_\phi^2)+\cos^4\theta(a^4\frac{E^2}{c^8}v^2v_\theta-a^2\frac{E}{c^4}r^2v_\theta v_\phi^2)\right]\frac{\partial^2\mathcal{G}}{\partial v\partial v_\theta}\right.\\
    &\left.+\frac{8}{v}\left[\cos^2\theta(a^2\frac{E}{c^4}r^2v_\theta^2v_\phi+r^4v_\theta^2v_\phi)+\cos^4\theta(a^4\frac{E^2}{c^8}v^2v_\phi-a^2\frac{E}{c^4}r^2v^2v_\phi-a^2\frac{E}{c^4}r^2v_\phi^3+r^4v_\phi^3)\right]\frac{\partial^2\mathcal{G}}{\partial v\partial v_\phi}\right.\\
    &\left.+\left[\cos^2\theta(-8a^2\frac{E}{c^4}r^2v_\theta v_\phi+8r^4v_\theta v_\phi)+\cos^4\theta(8a^4\frac{E^2}{c^8}v_\theta v_\phi-8a^2\frac{E}{c^4}r^2v_\theta v_\phi)\right]\frac{\partial^2\mathcal{G}}{\partial v_\theta\partial v_\phi}\right.\\
    &\left.+\frac{4r^4}{v^3}\left[v^2v_\theta^2-v_\theta^4-2v_\theta^2 v_\phi^2\cos^2\theta+\cos^4\theta(v^2v_\phi^2-v_\phi^4)\right]\frac{\partial\mathcal{G}}{\partial v}\right\},
\end{aligned}
\label{appendix eq: general LDC}
\end{equation}
and the cross terms:
\begin{equation}
    \begin{aligned}
        \langle\Delta E\Delta L_z\rangle&=\Gamma'\left[L_z\frac{\partial^2\mathcal{G}}{\partial v^2}+L_z\frac{\partial^2\mathcal{G}}{\partial v_\phi^2}+L_z\frac{v_\theta}{v}\frac{\partial^2\mathcal{G}}{\partial v\partial v_\theta}+L_z\frac{v^2+v_\phi^2}{vv_\phi}\frac{\partial^2\mathcal{G}}{\partial v\partial v_\phi}+L_z\frac{v_\theta}{v_\phi}\frac{\partial^2\mathcal{G}}{\partial v_\theta\partial v_\phi}\right],\\
        \langle\Delta E\Delta Q\rangle&=\Gamma'\left\{\left[2r^2v_\theta^2+2(-a^2\frac{Ev^2}{c^4}+r^2v_\phi^2)\cos^2\theta\right]\frac{\partial^2\mathcal{G}}{\partial v^2}+2v_\theta^2(r^2-a^2\frac{E}{c^4}\cos^2\theta)\frac{\partial^2\mathcal{G}}{\partial v_\theta^2}+2v_\phi^2(r^2-a^2\frac{E}{c^4})\cos^2\theta\frac{\partial^2\mathcal{G}}{\partial v_\phi^2}\right.\\
        &\left.+\frac{2v_\theta}{v}\left[r^2v^2+r^2v_\theta^2+\cos^2\theta(r^2v_\phi^2-2a^2\frac{Ev^2}{c^4})\right]\frac{\partial^2\mathcal{G}}{\partial v\partial v_\theta}+\frac{2v_\phi}{v}\left[r^2v_\theta^2+\cos^2\theta(r^2v^2+r^2v_\phi^2-2a^2\frac{Ev^2}{c^4})\right]\frac{\partial^2\mathcal{G}}{\partial v\partial v_\phi}\right.\\
        &\left.+\left[2r^2v_\theta v_\phi+2v_\theta v_\phi\cos^2\theta(r^2-2a^2\frac{E}{c^4})\right]\frac{\partial^2\mathcal{G}}{\partial v_\theta\partial v_\phi}\right\},\\
        \langle\Delta L_z\Delta Q\rangle&=\Gamma'\left\{\frac{2L_z}{v^2}\left[r^2v_\theta^2+\cos^2\theta(r^2v_\phi^2-a^2\frac{Ev^2}{c^4})\right]\frac{\partial^2\mathcal{G}}{\partial v^2}+2L_z\cos^2\theta(r^2-a^2\frac{E}{c^4})\frac{\partial^2\mathcal{G}}{\partial v_\phi^2}\right.\\
        &\left.+\frac{2L_zv_\theta}{v}(r^2-a^2\frac{E}{c^4}\cos^2\theta)\frac{\partial^2\mathcal{G}}{\partial v\partial v_\theta}+\frac{2L_z}{vv_\phi}\left[r^2v_\theta^2-\cos^2\theta(-2r^2v_\phi^2+a^2\frac{E}{c^4}(v^2+v_\phi^2))\right]\frac{\partial^2\mathcal{G}}{\partial v\partial v_\phi}\right.\\
        &\left.+\frac{2L_zv_\theta}{v_\phi}(r^2-a^2\frac{E}{c^4}\cos^2\theta)\frac{\partial^2\mathcal{G}}{\partial v_\theta\partial v_\phi}-\frac{2L_zr^2}{v^3}\left[v_\theta^2+(v_\phi^2-v^2)\cos^2\theta\right]\frac{\partial\mathcal{G}}{\partial v}\right\}.
    \end{aligned}
    \label{appendix eq: general LDC cross terms}
\end{equation}

\begin{figure}
    \includegraphics[width=85mm]{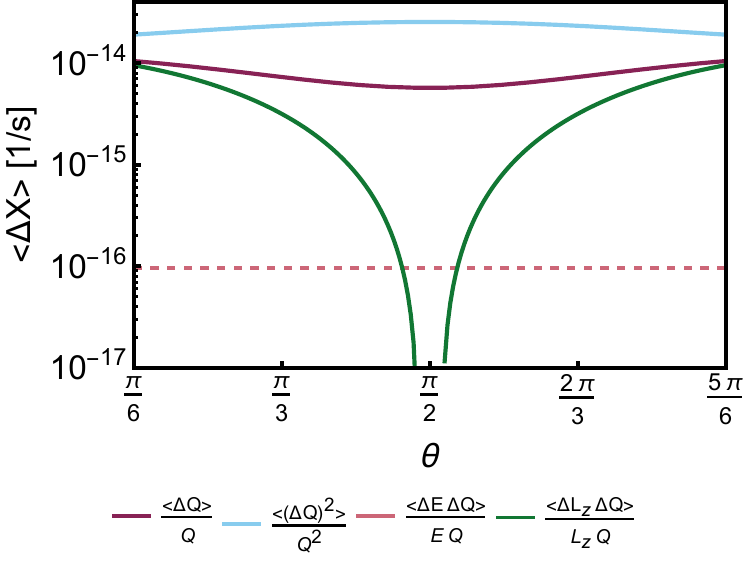}
    \includegraphics[width=85mm]{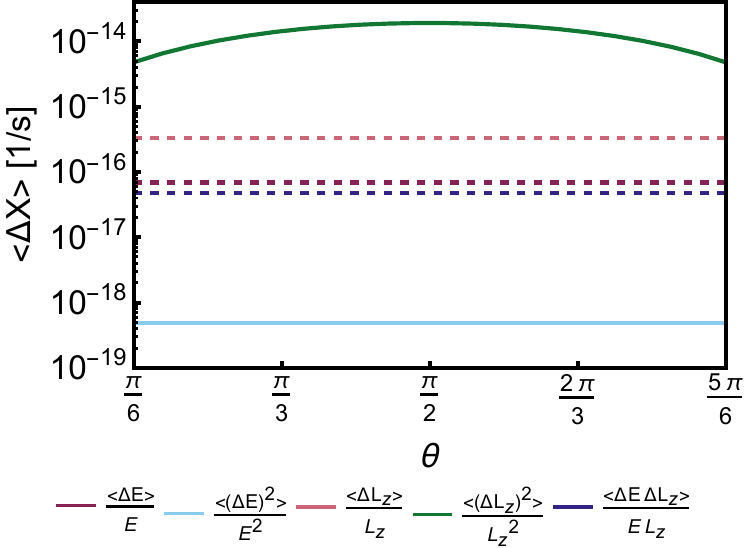}\\
    \includegraphics[width=85mm]{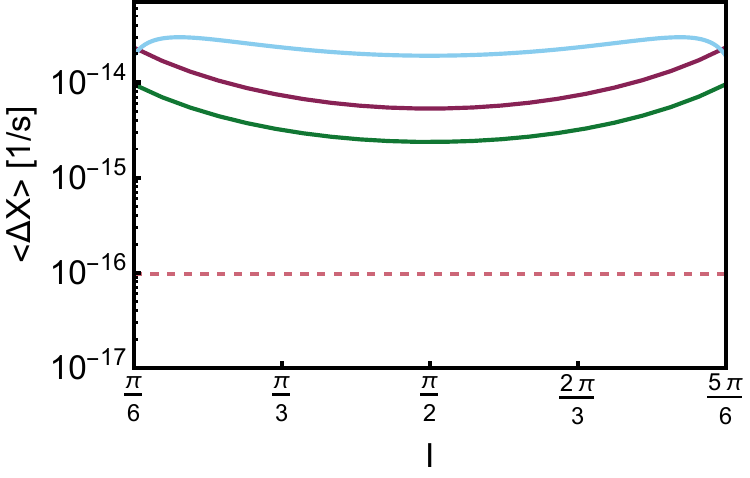}
    \includegraphics[width=85mm]{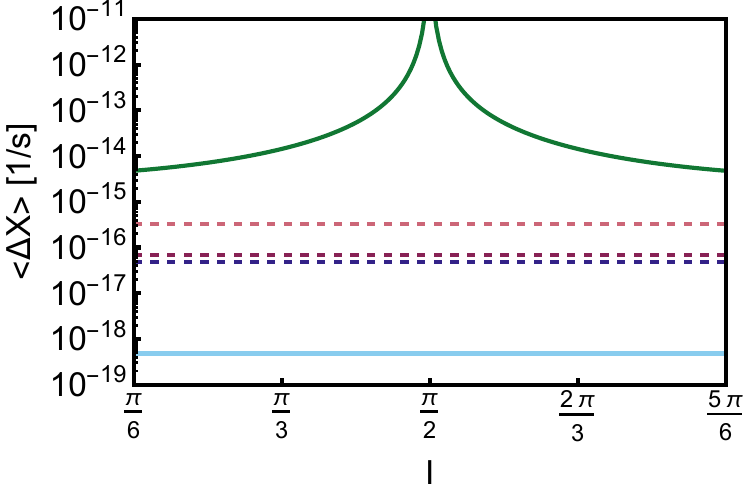}
    \caption{A comparison of local diffusion coefficients as a function of colatitude, $\theta$ with $I=60\degree$ (top), or inclination with $\theta=60\degree$ (bottom). The new diffusion coefficients are in the left panels, while the previously used coefficients are in the right panels.  
    The semimajor axis $a=10^7 r_{\rm g}$, eccentricity $e=0.99$, and the radius is taken at the apocenter. The mass of the MBH is $M_\bullet=10^6M_\odot$, and the mass of the CO is $m_\bullet=10M_\odot$. 
    }  
    \label{fig: LDC}
    
\end{figure}

The local diffusion coefficients are presented in \cref{fig: LDC}

\section{Alternate GDCs}
\label{appendix sec: alternative GDC}

Here we provide alternative expressions for the diffusion coefficients of the integrals of motion, using the simplified assumptions introduced in \S\ref{sec:local}. In this approximation, the potential $\Phi(r)$ depends only on the radial coordinate, the field-star velocity distribution is isotropic, and two-body scatterings are assumed to dominate at the apocenter, where the contribution of the MBH spin is negligible.

\subsection{$\cot{\iota}$}
Instead of evolving $L_z$, one may evolve the inclination $\iota$ directly. Here we present the diffusion coefficients of $\cot \iota = L_z / \sqrt{Q}$, which is a continuous and unbounded parameter.

\begin{equation}
\begin{aligned}
\langle\Delta \cot\iota\rangle&=-\frac{1}{2}\frac{L_z}{Q^{3/2}}\langle\Delta Q\rangle+\frac{3}{8}\frac{L_z}{Q^{5/2}}\langle(\Delta Q)^2\rangle
+\frac{1}{\sqrt{Q}}\langle\Delta L_z\rangle-\frac{1}{2}\frac{1}{Q^{3/2}}\langle\Delta L_z\Delta Q\rangle\\
&=\frac{r^2L_z}{4Q^{5/2}}\left[2Q-3\cos^2\theta(Q+L_z^2)\right]\langle(\Delta v_\perp)^2\rangle,\\
\langle(\Delta \cot\iota)^2\rangle&=\frac{1}{Q}\langle(\Delta L_z)^2\rangle+\frac{1}{4}\frac{L_z^2}{Q^3}\langle(\Delta Q)^2\rangle-\frac{L_z}{Q^2}\langle\Delta L_z\Delta Q\rangle\\
&=\frac{r^2}{2Q^3}\left(Q+L_z^2\right)\left[Q-\cos^2\theta\left(Q+L_z^2\right)\right]\langle(\Delta v_\perp)^2\rangle,\\
\langle\Delta E\Delta\cot\iota\rangle&=0,\\
\langle\Delta L_z\Delta\cot\iota\rangle&=\frac{1}{\sqrt{Q}}\langle(\Delta L_z)^2\rangle-\frac{L_z}{2Q^{3/2}}\langle\Delta L_z\Delta Q\rangle\\
&=\frac{r^2\sin^2\theta}{2Q^{3/2}}(Q-L_z^2\cot^2\theta)\langle(\Delta v_\perp)^2\rangle.
\end{aligned}
\end{equation}

\subsection{Effective angular momentum}
We can define an effective angular momentum squared as $K\equiv L_z^2+Q$. With this definition, $K$ must satisfy $K\geq Q$ and $K\geq L_z^2$. Using the set $(E, K, \cot \iota)$ as the three integrals of motion removes the need to impose these boundary conditions.

\begin{equation}
\begin{aligned}
\langle\Delta K\rangle&=\langle\Delta Q\rangle+2L_z\langle\Delta L_z\rangle+\langle(\Delta L_z^2)\rangle\\
&=\frac{2K^2}{v}\langle\Delta v_\parallel\rangle+\frac{K^2}{v^2}\langle(\Delta v_\parallel)^2\rangle+\frac{1}{2v^2}(2r^2v^2-K^2)\langle(\Delta v_\perp)^2\rangle\\
\langle(\Delta K)^2\rangle&=4L_z^2\langle(\Delta L_z)^2\rangle+\langle(\Delta Q)^2\rangle+4L_z\langle\Delta L_z\Delta Q\rangle\\
&=\frac{4K^2}{v^2}\langle(\Delta v_\parallel)^2\rangle+\frac{2K}{v^2}(r^2v^2-K)\langle(\Delta v_\perp)^2\rangle\\
\langle\Delta E\Delta K\rangle&=2L_z\langle\Delta E\Delta L_z\rangle+ \langle\Delta E\Delta Q\rangle\\
&=2K\langle(\Delta v_\parallel)^2\rangle\\
\langle\Delta L_z\Delta K\rangle&=2L_z\langle(\Delta L_z)^2\rangle+\langle\Delta L_z\Delta Q\rangle\\
&=\frac{2L_zK}{v^2}\langle(\Delta v_\parallel)^2\rangle+\frac{L_z}{v^2}(r^2v^2-K)\langle(\Delta v_\perp)^2\rangle\\
\langle\Delta K\Delta \cot\iota\rangle&=0
\end{aligned}
\end{equation}

\subsection{Square root}
Here we provide a general expression for taking the square root of a diffusion coefficient of a constant of motion $X$: 
\begin{equation}
\begin{aligned}
    \langle\Delta \sqrt{X}\rangle&=\frac{1}{2\sqrt{X}}\langle\Delta X\rangle-\frac{1}{8X^{3/2}}\langle(\Delta X)^2\rangle\\
    \langle(\Delta \sqrt{X})^2\rangle&=\frac{1}{4X}\langle(\Delta X)^2\rangle\\
    \langle\Delta \sqrt{X} \Delta Y\rangle&=\frac{1}{2\sqrt{X}}\langle\Delta X\Delta Y\rangle
\end{aligned}
\end{equation}

\subsection{Natural logarithm}
It can be useful to take the natural logarithm of a positive-definite constant of motion so that it becomes defined over $(-\infty,\infty)$. Below is a general expression for transforming the diffusion coefficient of a constant of motion $X$ to $\ln X$:

\begin{equation}
\begin{aligned}
    \langle\Delta \ln{X}\rangle&=\frac{1}{X}\langle\Delta X\rangle-\frac{1}{2X^2}\langle(\Delta X)^2\rangle\\
    \langle(\Delta \ln{X})^2\rangle&=\frac{1}{X^2}\langle(\Delta X)^2\rangle\\
    \langle\Delta \ln{X} \Delta Y\rangle&=\frac{1}{X}\langle\Delta X\Delta Y\rangle
\end{aligned}
\end{equation}

\section{Orbit Averaged Diffusion Coefficients}
\label{appendix sec: orbit average}
The orbit-averaged diffusion coefficient, $\langle\Delta V^i\rangle$ where $V^i\in \{E,L_z,Q\}$, is given by
\begin{equation}
    \langle\Delta V^i\rangle_{\rm o}\equiv\frac{\int_{\rm orb}\langle\Delta V^i\rangle a^{1/2}\left|\frac{\partial(v,v_\theta,v_\phi)}{\partial(E,L_z,Q)}\right|d\Vec{x}}{\int_{\rm orb}a^{1/2}\left|\frac{\partial(v,v_\theta,v_\phi)}{\partial(E,L_z,Q)}\right|d\Vec{x}}
    \label{appendix eq: general orbit avg}
\end{equation}
where $a$ denotes the determinant of the metric from \cref{appendix eq: metric}, and $\left|\frac{\partial(v,v_\theta,v_\phi)}{\partial(E,L_z,Q)}\right|$ is the Jacobian determinant of the coordinate transformation from $(q^1=v,q^2=v_\theta,q^3=v_\phi)$ to $(E,L_z,Q)$. 
Calculating the determinant of the metric and the Jacobian, we obtain:
\begin{equation}
\begin{aligned}
a^{1/2}\left|\frac{\partial(v,v_\theta,v_\phi)}{\partial(E,L_z,Q)}\right|&=\frac{v}{v_r}\frac{1}{2r^3vv_\theta\sin\theta}
=\frac{1}{2r^3v_r v_\theta\sin\theta}.
\end{aligned}
\label{appendix eq: Jacobian}
\end{equation}

Since the local diffusion coefficients depend on $r$ and $\theta$, the integration is:
\begin{equation}
 \int_{\rm orb}d\Vec{x}=2\int_{\theta_{\rm min}}^{\theta_{\rm max}}\int_{r_{\rm p}}^{r_{\rm a}}2\pi r^2\sin\theta {\rm d}r{\rm d}\theta   
 \label{appendix eq: avg integrator}
\end{equation}
where $r_{\rm p}$ and $r_{\rm a}$ are the pericenter and apocenter, respectively, and 
$\theta_{\rm min}=\pm(\pi/2-I)$ and  $\theta_{\rm max}=\pi-\theta_{\rm min}$ are the minimum and maximum values of $\theta$ along the orbit for a given inclination $I$, where the upper sign refers to prograde orbits, while the lower sign refers to retrograde orbits.

Combining \cref{appendix eq: general orbit avg,appendix eq: Jacobian,appendix eq: avg integrator}, the orbit averages are:
\begin{equation}
    \langle\Delta V^i\rangle_{\rm o}\equiv\frac{\int_{\theta_{\rm min}}^{\theta_{\rm max}}\int_{r_{\rm p}}^{r_{\rm a}} J\langle\Delta V^i\rangle {\rm d}r{\rm d}\theta}{\int_{\theta_{\rm min}}^{\theta_{\rm max}}\int_{r_{\rm p}}^{r_{\rm a}} J{\rm d}r{\rm d}\theta}
\end{equation}
where $J=\frac{1}{rv_rv_\theta}$. 

When calculating the local diffusion coefficients, we assume that they are dominant near apocenter, far from the MBH. However, when performing the orbital average, we need to be more careful, since along the orbit the test particle approaches the MBH near pericenter, where relativistic effects can become important. For instance, the orbit undergoes precession, so $\theta_{\rm min}$ and $\theta_{\rm max}$ do not coincide with pericenter and apocenter. Consequently, the averaging should be carried out over geodesics rather than Keplerian orbits. In particular, the squared velocity $v^2=v_r^2+v_\theta^2+v_\phi^2$ is computed from the geodesic equations \citep{Bardeen+1972} where:
\begin{equation}
\begin{aligned}
    v_r^2&=\dot{r}^2=R(r)/\Sigma(r,\theta)^2\\
    v_\theta^2&=r^2\dot{\theta}^2=r^2\Theta(\theta)/\Sigma(r,\theta)^2\\
v_\phi^2&=r^2\sin^2\theta\dot{\phi}^2=r^2\sin^2\theta\Phi(r, \theta)/\Sigma(r,\theta)^2
\end{aligned}
\end{equation}
where the derivatives are with respect to proper time, $R(r)$, $\Theta(\theta)$ and $\Phi(r,\theta)$ are the radial, polar and azimuthal potentials of the Kerr metric, and $\Sigma=r^2+\chi_\bullet^2\cos^2\theta$. The integrand prefactor is therefore $J=\frac{\Sigma(r,\theta)^2}{r^2 \sqrt{R(r)}\sqrt{\Theta(\theta)}}$. Also, we note that the integration boundaries are the roots of the respective potentials.

\blue{Because the four-velocity components are defined with respect to proper time $\tau$, the integration over $J\mathrm{d}r\mathrm{d}\theta \propto \mathrm{d}\tau$ yields an average over proper time rather than coordinate time $t$. Formally, the Fokker-Planck equation evolves relative to the coordinate time $t$, which requires an additional weighting factor of $\mathrm{d}t/\mathrm{d}\tau = u^t$ in the orbital integral. However, because scatterings occur mostly near apocenter, where $\mathrm{d}t/\mathrm{d}\tau \approx 1$, the numerical difference between proper-time and coordinate-time averages is negligible.}

Additionally, we tested the accuracy of using the Newtonian form of this Jacobian:  
$J=\left[\left(2(E-c^2-\Phi(r))-L_z^2/r^2-Q/r^2\right)\left(Q-L_z^2\cot^2\theta\right)\right]^{-1/2}$
and found that the deviation from the geodesic result is small.

\section{Gravitational waves Losses}
\label{appendix sec: GWs}

In this paper, we compute the GW fluxes using the ``kludge'' method described in \cite{Gair&Glampedakis2006}. \blue{For $E$ we use Eq. 20, for $L_z$ we use Eq. 59, and for $Q$ we use Eq. 60.} To check the impact of the GWs on the orbital inclination, we compute the orbital evolution driven solely by GW emission. In \cref{appendix fig: inclination rp}, we present the evolution of the orbital inclination and find that it changes only weakly over the course of the inspiral. This indicates that the inclination evolution calculated in this work is primarily driven by inclination diffusion due to two-body scatterings rather than by GW emission. In addition, we compare the approach of \cite{Gair&Glampedakis2006} with that of \cite{Peters1964}. We find in \cref{appendix fig: GG vs Peters} that the differences between the two methods become relevant only close to the MBH, where GW emission has already circularized the orbit.

\begin{figure}
    \centering
    \includegraphics[width=85mm]{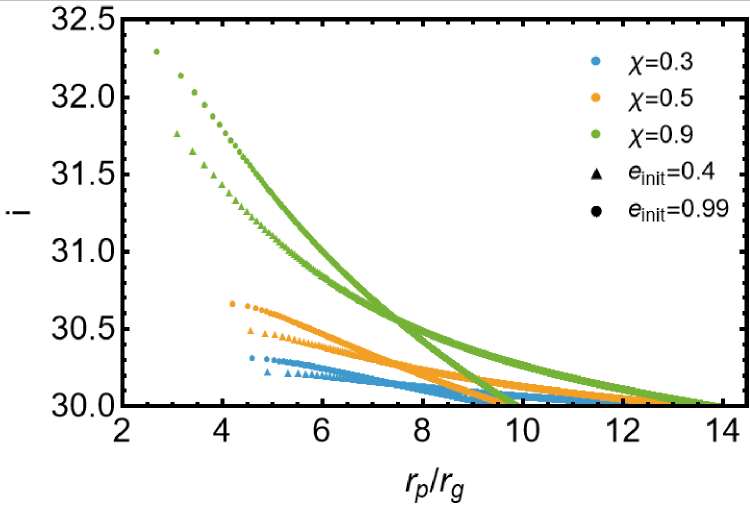}
    \includegraphics[width=85mm]{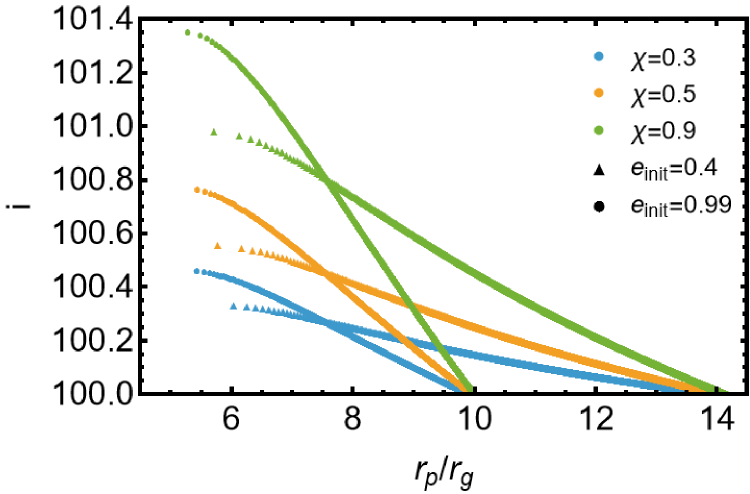}
    \caption{Inclination evolution as a function of pericenter radius due to GW emission calculated using \cite{Gair&Glampedakis2006}, for different MBH spins and initial orbital parameters. We can see only a small change in the orbital inclination.}
    \label{appendix fig: inclination rp}
\end{figure}

\begin{figure}
    \centering
    \includegraphics[width=85mm]{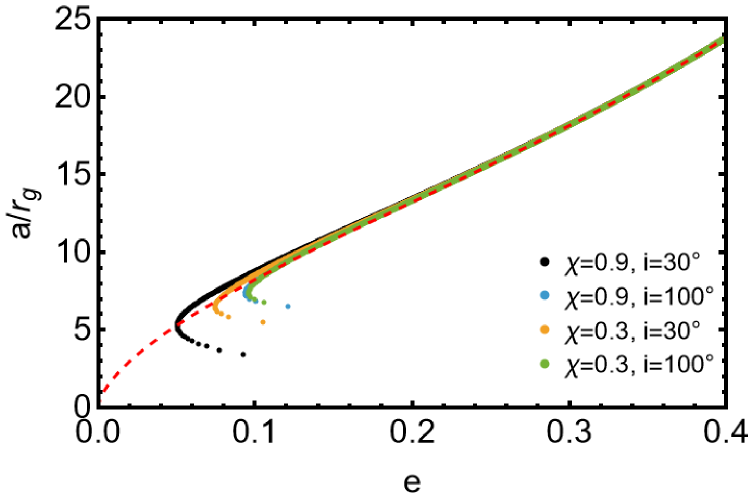}
    \includegraphics[width=85mm]{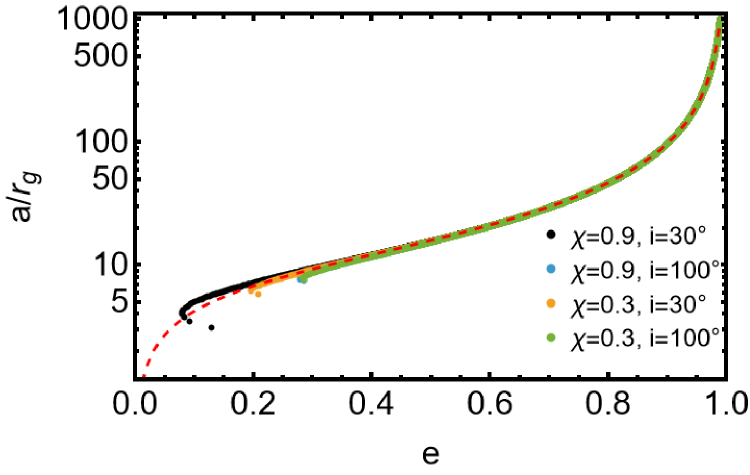}
    \caption{Comparisons between the semimajor axis as a function of eccentricity between \cite{Gair&Glampedakis2006} (colored solid curves) and \cite{Peters1964} (red dashed curve). The comparisons are presented for different MBH spins, and initial semimajor axes, eccentricities and inclinations. }
    \label{appendix fig: GG vs Peters}
\end{figure}

\end{document}